\documentclass[english,superscriptaddress,floatfix,twocolumn,oneside, amsmath,amssymb,amsfonts,aps,pra,floatfix]{revtex4-1}
\usepackage[english]{babel}
\usepackage[utf8x]{inputenc}
\usepackage{microtype}
\usepackage{graphicx}
\usepackage{siunitx}
\usepackage{xspace}
\usepackage{xcolor}
\usepackage[hidelinks]{hyperref}
\usepackage{bm}
\usepackage{bbold}
\usepackage{mathrsfs} 
\usepackage{natbib}
\usepackage{csquotes}
\usepackage{physics}
\usepackage{ifthen}
\usepackage{qcircuit}
\usepackage{algpseudocode}
\usepackage{algorithm}

\usepackage[draft]{fixme}

\usepackage[capitalise]{cleveref}

\newcommand{\affA}{Department of Physics and Astronomy, Aarhus University, DK-8000 Aarhus C, Denmark}
\newcommand{\affB}{Aarhus Institute of Advanced Studies, Aarhus University, DK-8000 Aarhus C, Denmark}

\newcommand{\cnot}{\textsc{cnot}\xspace}
\newcommand{\cns}{\textsc{cns}\xspace}
\newcommand{\zz}{\textsc{zz}\xspace}
\newcommand{\x}{\textsc{x}\xspace}
\newcommand{\xx}{\textsc{xx}\xspace}
\newcommand{\z}{\textsc{z}\xspace}
\newcommand{\y}{\textsc{y}\xspace}
\newcommand{\h}{\textsc{h}\xspace}
\renewcommand{\r}{\textsc{r}\xspace}
\newcommand{\Cr}{\textsc{cr}\xspace}
\newcommand{\ii}{\textsc{ii}\xspace}
\newcommand{\cz}[1][]{\ensuremath{\textsc{c}^{#1}\textsc{z}}\xspace}

\newcommand{\cswap}[1][]{%
	\ifthenelse{\equal{#1}{}}{\textsc{swap}\xspace}{\ifthenelse{\equal{#1}{1}}{\textsc{cswap}\xspace}{\ensuremath{\textsc{c}^{#1}\textsc{swap}\xspace}}}}

\newcommand{\ciswap}[1][]{%
	\ifthenelse{\equal{#1}{}}{\ensuremath{ i\textsc{swap}}\xspace}{\ifthenelse{\equal{#1}{1}}{\ensuremath{ \textsc{c} i \textsc{swap}\xspace}}{\ensuremath{\textsc{c}^{#1}i\textsc{swap}}\xspace}}}

\date{\today}
\begin{document}
	
	\title{Application of the Diamond Gate in Quantum Fourier Transformations and Quantum Machine Learning}
	
	\author{E. Bahnsen}
	\affiliation{\affA}
	\author{S. E. Rasmussen}
	\affiliation{\affA}
	\author{N. J. S. Loft}
	\affiliation{\affA}
	\author{N. T. Zinner}
	\email{zinner@phys.au.dk}
	\affiliation{\affA}
	\affiliation{\affB}
	
	\begin{abstract}
		As we are approaching actual application of quantum technology, it is essential to exploit the current quantum resources in the best possible way. With this in mind, it might not be beneficial to use the usual standard gate sets, inspired from classical logic gates, when compiling quantum algorithms when other less standardized gates currently perform better. We, therefore, consider a promising native gate, which occurs naturally in superconducting circuits, known as the diamond gate. We show how the diamond gate can be decomposed into standard gates and, using single-qubit gates, can work as a controlled-not-swap (\cns) gate. We then show how this \cns gate can create a controlled phase gate. Controlled phase gates are the backbone of the quantum Fourier transform algorithm, and we, therefore, show how to use the diamond gate to perform this algorithm. We also show how to use the diamond gate in quantum machine learning; namely, we use it to approximate non-linear functions and classify two-dimensional data.
	\end{abstract}
	
	\maketitle
	
	\section{Introduction}\label{sec:intro}
	
	Quantum technology has gained an increasing amount of attention in the past couple of years, especially since the quantum supremacy experiment at Google \cite{Arute2019}. The search for quantum advantages in practical applications is an increasingly active area of research \cite{Bravyi2020}, and there is a growing consensus that practical applications will be found in the noisy intermediate-scale quantum (NISQ) era \cite{Preskill2018}. In this era, quantum technology supports only a couple of tens of qubits and a few hundred gate operations before the noise becomes too overwhelming.
	
	In this era, we must work with the native gates of quantum systems, despite their imperfections compared to the set of standard gates, composed of gates such as the \cnot or Fredkin gate. An example of such a native gate is the diamond gate \cite{Loft2020}, which has the advantage of being a naturally arising highly entangling four-qubit gate in superconducting circuits. However, it has the disadvantage of being a complex gate to understand and thus use in quantum algorithm circuits. This paper seeks to interpret the gate and show its various applications in synthesizing other quantum gates, performing quantum algorithms, and participating in quantum machine learning. 
	This paper aims to show the utility of the diamond gate and show a different approach to quantum circuitry using some currently available gates rather than some of the standard gates.
	
	The diamond gate is closely related to the Fredkin gate, which is often used in a set of standard gates for quantum computation \cite{Nielsen2010}. In particular, they are both controlled-swapping gates; however, the diamond has a pair of control qubits, which must be in a Bell state for the closed state of the gate. Other implementations of controlled-swapping gates besides the Fredkin gate \cite{Milburn1989,Chau1995,Fiuraifmmode2006,Fiuraifmmode2008,Gong2008,Patel2016,Ono2017,Smolin1996,Baekkegaard2019} include superconducting qubits in a waveguide cavity \cite{Poletto2012}, linear superconducting qubits \cite{Rasmussen2019}, and multicontrolled \ciswap gates \cite{Rasmussen2020a}.
	
	A popular approach to optimize the computational power is to divide the computational task between classical and quantum resources, i.e., a so-called hybrid quantum classical (HQC) algorithm. Examples of such HQC algorithms are the quantum approximate optimization algorithm (QAOA) \cite{Fahri2014,Otterbach2017,Moll2018}, the quantum autoencoder (QAE) \cite{Romero2017}, the quantum variational error corrector (QVECTOR) \cite{Johnson2017}, classification via the near term-quantum neural network (QNN) \cite{Fahri2018,Havlicek2019,Schuld2020,Hubregtsen2020}, the quantum generative adversarial network (QuGAN) \cite{Dallaire-Demers2018,Lloyd2018,Zoufal2019,Zhu2019}, and the variational quantum eigensolver (VQE) \cite{Peruzzo2014,McClean2016,Omalley2016,Kandala2017,Cao2019,Barkoutos2018,McCaskey2019,Gard2020}.
	
	Common to all these HQC algorithms is that they share a subroutine for producing parameterized trial states, where the parameters can be tuned to optimize a function value. The performance of these algorithms depends on the configuration of the parameterized quantum circuit (PQC) \cite{Sim2019,Geller2018,Du2018,Benedetti2019}. Many of the HQC algorithms draw inspiration from classical machine-learning algorithms. Therefore, PQCs are often considered the quantum-mechanical equivalent of neural networks, which is why they are sometimes referred to as quantum neural networks \cite{Fahri2018,Havlicek2019,Schuld2020,Hubregtsen2020}.
	In this paper, we apply the diamond gate to quantum circuit learning \cite{Mitarai2018} and classification via QNNs \cite{Hubregtsen2020}, using the diamond gate as an entangling gate in a PQC. Previously, it has been shown that the diamond gate could be useful in VQE algorithms \cite{Rasmussen2020c}. Besides HQC algorithms, we also consider the diamond gate for performing quantum Fourier transformations (QFT), which are used to estimate the phase of quantum-mechanical amplitudes. This could be used to solve the order-finding problem or the factoring problem \cite{Nielsen2010}.	
	
	The rest of the paper is structured as follows. In \cref{sec:diamond}, we introduce the diamond gate and its main properties. In \cref{sec:cnsEquivalent}, we show how to change the diamond gate into a combined \cswap and \cnot gate, also known as the \cns gate. Then, in \cref{sec:decomposition}, we symmetrically decompose the diamond gate into standard gates while preserving its symmetry. Using this decomposition, we show in \cref{sec:nativePhaseGate} that the diamond gate can be operated as a native controlled-phase gate. We then, in \cref{sec:QFT}, use this to perform the quantum Fourier transform with the diamond gate. In \cref{eq:MLwDiamond} we apply the diamond gate to machine-learning problems. In particular, we apply it to a problem of quantum circuit learning \cite{Mitarai2018} and classification of two-dimensional data \cite{Hubregtsen2020}. Finally, in \cref{sec:conclusion}, we present our conclusion and the outlook.

	\section{The four-qubit diamond gate}\label{sec:diamond}
	
	The algorithms presented in this paper are based upon the diamond gate developed in Ref. \cite{Loft2020}; therefore, we only offer an overview and details can be found in the original paper.
	
	\begin{figure}
		\centering
		\includegraphics{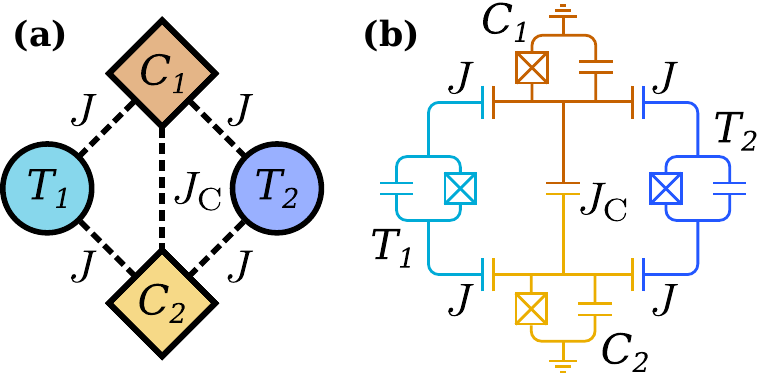}
		\caption{\textbf{(a)} A four-qubit system of two control qubits ($C_1$ and $C_2$) and two target qubits ($T_1$ and $T_2$), which constitutes the diamond circuit. These couple through their exchange interaction with strengths $J$ and $J_C$ as indicated.
		\textbf{(b)} An equivalent simple circuit diagram of superconducting transmon qubits, which are capacitively coupled with the same strengths $J$ and $J_C$.
		Figure reproduced, with permission, from Ref. \cite{Loft2020}.}
		\label{fig:diamond}
	\end{figure}
	
	The diamond gate is a highly entangled four-qubit gate natively implemented using superconducting circuits. The system consists of two target qubits ($T_1$ and $T_2$) and two control qubits ($C_1$ and $C_2$). Each pair of control and target qubits is connected with a coupling strength of $J$. The two control qubits are coupled with a coupling strength of $J_C$. The system schematics can be seen in \cref{fig:diamond}(a). A superconducting circuit design resulting in the diamond gate could consist of transmonlike qubits \cite{Koch2007,Schreier2008} coupled by capacitors, as seen in \cref{fig:diamond}(b). 
	
	Using units where $\hbar = 1$, the system is governed by the Hamiltonian
	\begin{equation}
	\begin{aligned}
	H=&- \frac{1}{2}(\Omega + \Delta)(\sigma_z^{T_1} + \sigma_z^{T_2}) -\frac{1}{2}\Omega(\sigma_z^{C_1} + \sigma_z^{C_2})\\
	&+ J_C \sigma_y^{C_1} \sigma_y^{C_2} + J(\sigma_y^{T_1} + \sigma_y^{T_2})(\sigma_y^{C_1} + \sigma_y^{C_2}),
	\end{aligned}
	\end{equation}
	where $\sigma_{x,y,z}^j$ are the Pauli operators acting on the $j$th qubit. The target qubits are in resonance with each other, with frequency $\Omega+\Delta$, while the control qubits are in resonance with frequency $\Omega$.
	
	Assuming that $|2\Omega| \gg |J|$ and $|\Delta| \gg |J|, |J_C|$ the time-evolution operator of the diamond circuit takes the form
	\begin{equation}\label{eq:Ut}
	\begin{aligned}
	U(t) &=
	\ketbra{00}{00}\otimes U_T^{00}(t) +
	\ketbra{11}{11}\otimes U_T^{11}(t) \\&+
	\ketbra{\Psi^+}{\Psi^+}\otimes U_T^{\Psi^+}(t) +
	\ketbra{\Psi^-}{\Psi^-}\otimes U_T^{\Psi^-}(t),
	\end{aligned}
	\end{equation}
	where the unitary $U^i_T$ act on the target qubits, while the operators, $\ketbra{C_1C_2}{C_1C_2}$, acts on the control qubits, with $C_1$ and $C_2$ representing the first and second control qubit, respectively. These can be in either of the control states $\ket{00}, \ket{11}, \ket{\Psi^+}, \ket{\Psi^-}$, where $\ket{\Psi^\pm} = (\ket{01} \pm \ket{01})/\sqrt{2}$ are the Bell states. From \cref{eq:Ut}, we realize that $U(t)$ is block diagonal in this basis, which we call the Bell basis. The unitaries of the target qubits are (represented in the computational basis $\ket{T_1T_2} \in \{\ket{00}, \ket{01}, \ket {10}, \ket{11}\}$, which is used unless noted otherwise)
	\begin{subequations}\label{eq:U_T_operators}
		\begin{align}
		&U_T^{00}(t)
		=
		\begin{pmatrix}
		1 & 0 & 0 & 0 \\
		0 & \frac{1}{2}(e^{-i\zeta t}+1) & \frac{1}{2}(e^{-i\zeta t}-1) & 0 \\
		0 & \frac{1}{2}(e^{-i\zeta t}-1) & \frac{1}{2}(e^{-i\zeta t}+1) & 0 \\
		0 & 0 & 0 & e^{-i\zeta t}
		\end{pmatrix},\\
		&U_T^{11}(t)
		=
		\begin{pmatrix}
		e^{i\zeta t} & 0 & 0 & 0 \\
		0 & \frac{1}{2}(e^{i\zeta t}+1) & \frac{1}{2}(e^{i\zeta t}-1) & 0 \\
		0 & \frac{1}{2}(e^{i\zeta t}-1) & \frac{1}{2}(e^{i\zeta t}+1) & 0 \\
		0 & 0 & 0 & 1
		\end{pmatrix},\\
		&U_T^{\Psi^+}(t)
		=
		\begin{pmatrix}
		e^{i\zeta t} & 0 & 0 & 0 \\
		0 & 1 & 0 & 0 \\
		0 & 0 & 1 & 0 \\
		0 & 0 & 0 & e^{-i\zeta t}
		\end{pmatrix}e^{-iJ_Ct},\\
		&U_T^{\Psi^-}(t)
		=
		\begin{pmatrix}
		1 & 0 & 0 & 0 \\
		0 & 1 & 0 & 0 \\
		0 & 0 & 1 & 0 \\
		0 & 0 & 0 & 1
		\end{pmatrix}e^{+iJ_Ct},
		\end{align}
	\end{subequations}
	where $\zeta = 4J^2/\Delta$.
	If we consider a gate time of $t_g = \pi/\zeta$, the unitaries of the target qubits can be written in standard gates as 
	\begin{subequations}\label{eq:UTg}
	\begin{align}
		&U^{00}_T(t_g) = \zz\times \cz \times \cswap,\label{eq:UTg00} \\
		&U^{11}_T(t_g) = - \cz \times \cswap, \\
		&U^{\Psi^+}_T(t_g) = - \zz\times e^{-iJ_C t_g}, \\
		&U^{\Psi^-}_T(t_g) = \ii \times e^{+iJ_C t_g},
	\end{align}
	\end{subequations}
	where we use the notation that $\zz = \z\otimes\z$ is a \z gate on each of the target qubits. Without external interactions and depending on the choice of coupling strengths, $J$ and $J_C$, the diamond gate can operate with fidelities above 0.99 and a gate time around \SI{60}{\nano \second}. Since it only contributes a phase, we set $J_C = 0$ for simplicity.

	\section{CNS equivalent circuit}\label{sec:cnsEquivalent}
	
	This section shows that the diamond gate, together with single-qubit rotations, can produce the useful \cns gate.
	If we consider the unitary of \cref{eq:UTg00} acting on the target qubits, we can construct the following circuit equivalent:
	\begin{equation*}
		\Qcircuit @C=0.3em @R=0.7em @!R {
			\lstick{\ket{0}} & \qw & \qw & \multigate{3}{U^{00}_T(t_g)} & \qw & \qw \\
			\lstick{\ket{0}} & \qw & \qw & \ghost{U^{00}_T(t_g)}	& \qw & \qw \\
			\lstick{T_1}     & \gate{\h} & \gate{\z} & \ghost{U^{00}_T(t_g)} & \qw & \qw & \\
			\lstick{T_2} & \qw & \gate{\z}& \ghost{U^{00}_T(t_g)} & \gate{\h} & \qw } \raisebox{-2.8em}{=}\hspace{0.75cm}
		\Qcircuit @C=1em @R=1em @!R {
			 \lstick{\ket{0}}&\qw  &\qw & \qw \\
			 \lstick{\ket{0}}&\qw  &\qw & \qw \\
			 \lstick{T_1} & \qswap & \ctrl{1} & \qw \\
			 \lstick{T_2} & \qswap\qwx & \targ{} & \qw }
	\end{equation*}
	from which we can create a combined \cswap and \cnot gate using the $U^{00}_T(t_g)$ unitary. The resulting gate is known as the \cns gate. This gate is, in its own right, helpful in implementing specific quantum circuits such as the Toffoli gate or, indeed, in quantum error-correction algorithms \cite{Schuch2003}.
	
	It is not possible to remove the swapping part of the above \cns gate using just the diamond gate or single-qubit rotations; however, it can be dealt with intuitively by swapping the remaining operations to its left.
	
	\section{Gate decomposition}\label{sec:decomposition}
	
	\begin{figure*}
		\centering
		\includegraphics{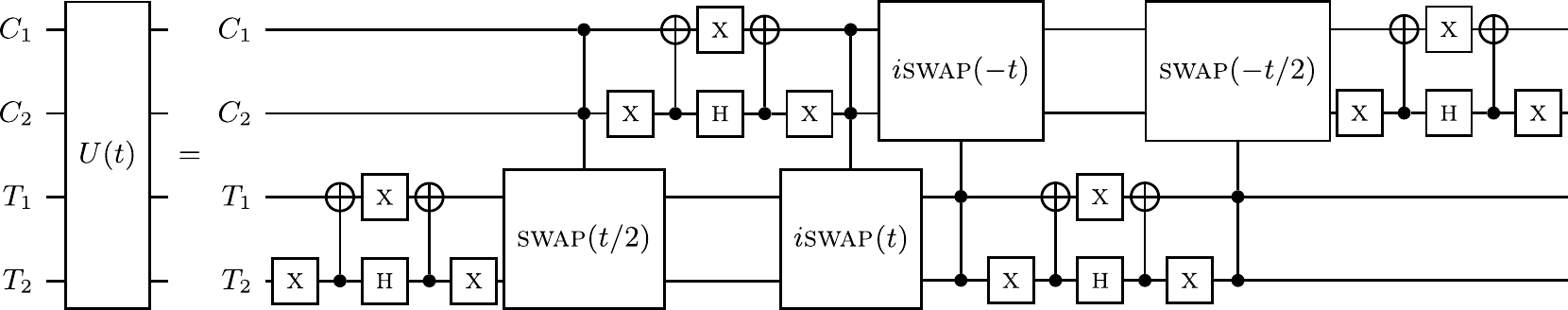}
		\caption{Standard gate decompositions of the diamond gate with no control coupling, $J_C = 0$, and application time $t$. The circuit employs two double-controlled parameterized \ciswap{} gates and two double-controlled parameterized \cswap{} gates.}
		\label{fig:U_circuit}
	\end{figure*}

	In Ref. \cite{Loft2020}, the diamond gate is decomposed into standard quantum circuit gates. In this case, the gate time is fixed at $t=t_g$ and the control qubit coupling, $J_C$, may be nonzero. The decomposition is done asymmetrically, focusing on the control qubits determining the transformation of the target qubits.
	
	Since we want to utilize the diamond gate for QFT, which requires controlled rotations on each qubit, we wish to create a symmetric decomposition in the sense that we want to preserve the symmetry between the control and target qubits. Therefore, we set $J_C=0$ and take the gate time as an arbitrary value. Considering \cref{fig:diamond}(a), we see that we can swap control and target qubits freely $C_1\leftrightarrow C_2$ and/or $T_1\leftrightarrow T_2$ and, when $J_C=0$, we can also swap the controls qubit with the target qubits $C_i \leftrightarrow T_i$ as long as we swap both sets of qubits at the same time; however, this swap also inverts time, $t\rightarrow -t$ (note that we can invert the time by $t\rightarrow 2t_g -t$). In other words, any choice of qubits is valid, as long as $C_1$ is opposite to $C_2$ and $T_1$ is opposite to $T_2$. Thus the superconducting diamond circuit has a fourfold rotational symmetry and a twofold mirror symmetry.
	
	With all of this symmetry in mind, we find that the diamond gate can be decomposed as in \cref{fig:U_circuit}. This circuit is relatively symmetric and consists of three parts that we discuss individually. The first part, consisting of a combination of \x gates and \cnot gates and the Hadamard gate, can be interpreted as an \x gate on each qubit in the Bell basis
	\begin{equation}\label{eq:xxInBell}
		\mathcal{V}^T\xx \mathcal{V}= \raisebox{1em}{
		\Qcircuit @C=0.6em @R=0.7em @!R {
			& \qw & \targ   & \gate{\x} & \targ   & \qw      & \qw \\ 
			& \gate{\x} & \ctrl{-1} & \gate{\h} & \ctrl{-1} & \gate{\x} & \qw      }}
	\end{equation}
	where $\xx = \x \otimes \x$ indicates an \x gate on each of the qubits and $\mathcal{V}$ is the transformation from the computational basis into the Bell basis. One can prove \cref{eq:xxInBell} using the fact that the transformation matrix is equivalent to the following circuit diagram:
	\begin{equation}
		\mathcal{V} = 
		\raisebox{1.1em}{\Qcircuit @C=0.6em @R=0.7em @!R { & \qw & \ctrl{1} & \ctrl{1} & \targ{}   & \gate{\x} & \qw\\
		 & \gate{\x} & \targ{1} & \gate{\h} & \ctrl{-1} & \gate{\x} & \qw }}
	\end{equation}
	This transformation is useful in regard to the diamond gate, since the diamond gate turns out to be diagonal in the Bell basis, a fact that is evident from \cref{eq:U_T_operators}.
	
	The second part of the circuit in \cref{fig:U_circuit} is a double-controlled parameterized \cswap gate, where the \cswap gate itself is
	\begin{equation}
		\begin{aligned}
		\cswap(t) &= \ketbra{00}{00} + \ketbra{11}{11} \\
		&+ (\ketbra{01}{01} + \ketbra{10}{10})\cos(\zeta t)\\
		&+ (\ketbra{01}{10} + \ketbra{10}{01})\sin(\zeta t),
		\end{aligned}
	\end{equation}
	which is the natural extension of the normal \cswap gate. We add two controls to this gate, making it a parameterized controlled Fredkin gate. Note that using the transformation to the Bell basis, the parameterized \cswap gate can be transformed into a controlled $\r_\z$ gate, something that is helpful in \cref{sec:nativePhaseGate}.
	
	The third part of the circuit in \cref{fig:U_circuit} is a double-controlled parameterized \ciswap gate, where the \ciswap gate itself is
	\begin{equation}
	\begin{aligned}
	\ciswap(t) &= \ketbra{00}{00} + \ketbra{11}{11} \\
		&+ \ketbra{\Psi^+}{\Psi^+}e^{-i\zeta t} + \ketbra{\Psi^-}{\Psi^-}e^{i\zeta t},
	\end{aligned}
	\end{equation}	
	which occurs naturally in many systems with XY interactions or Heisenberg models, such as solid-state systems \cite{Tanamoto2008,Tanamoto2009}, superconducting circuits \cite{You2005,Zagoskin2006,McKay2016,Dewes2012,Salathe2015,Vool2017,Krantz2019,Rasmussen2021}, and in cavity-mediated spin qubits and superconducting qubits \cite{Imamoglu1999,Benito2019,Blais2004}. Other implementations of the \ciswap gate are found in linear optics \cite{Wang2010,Bartkowiak2010} and nuclear spin using qudits \cite{Godfrin2018}. We consider a controlled version of this, with two control qubits \cite{Rasmussen2020b}, which is essentially a controlled Fredkin gate with a phase on the swapping part. 
	
	So far, we have assumed $J_C = 0$ but it is straightforward to add the functionality of $J_C\neq 0$ to the above circuit diagrams.
	This adds a global phase to the system if the control qubits are in the state $\ket{\Psi^-}$ or $\ket{\Psi^+}$.
	We exploit the transformation $\mathcal{V}$ from the computational basis to the Bell basis. This yields the circuit diagram seen in \cref{fig:add_JC} below. This circuit diagram can then be concatenated to the one in \cref{fig:U_circuit}.
	
	\begin{figure}
		\centering
		\includegraphics[scale=1]{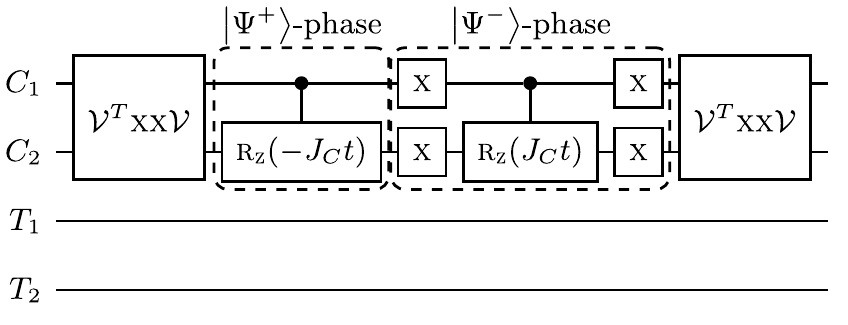}
		\caption{The Circuit diagram of how to add the functionality of the global phases that the diamond gate applies if $J_C\neq 0$.
		If the control qubits are in the state $\ket{\Psi^\pm}$, the phase $e^{\mp iJ_Ct}$ is applied.
		To add this functionality, we have to concatenate this to the end of the diagram of the diamond gate.}
		\label{fig:add_JC}
	\end{figure}
	
	\section{Native-gate derivatives}\label{sec:nativePhaseGate}
	
	\begin{figure}
		\centering
		\includegraphics[scale=0.6]{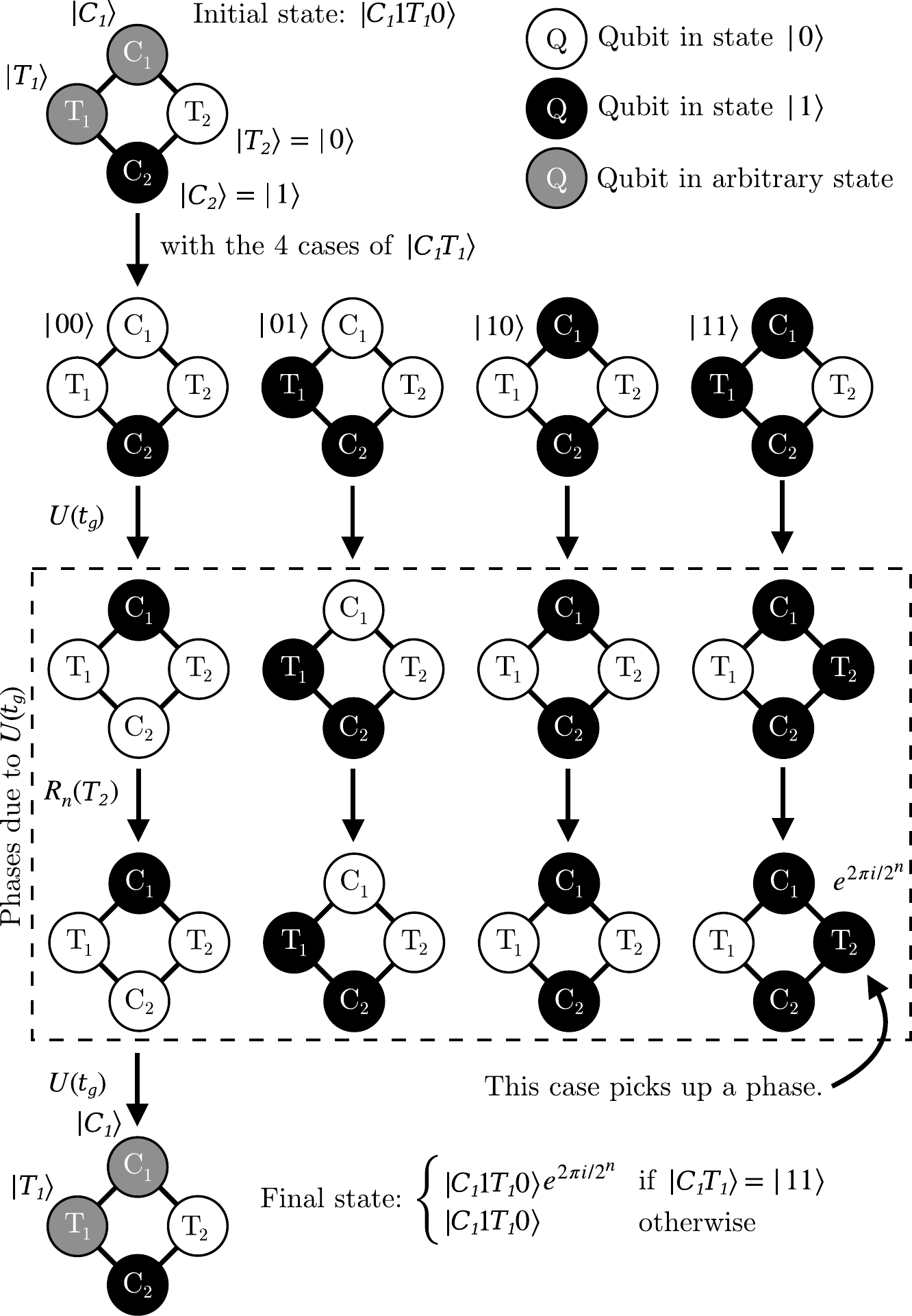}
		\caption{The flow chart of how the native diamond gate may implement a controlled $R_n$ gate. The qubits start in the state $\ket{C_11T_10}$ and will by the end pick up a phase $e^{2\pi i/2^n}$ if and only if both the control and the target qubit are in the $\ket{1}$ state. The flow chart corresponds to the circuit diagram in \cref{eq:Rn_diamond}.}
		\label{fig:URUflowChart}
	\end{figure}
	
	The decomposition of the diamond gate in \cref{fig:U_circuit} opens up for alternative interpretations of its functionality. We see that it might act as a controlled-\cswap{} or -phase gate.
	This would be particularly useful in the construction of a QFT circuit, which is highly dependent on controlled-phase gates; in particular, QFT is dependent on the implementation of the controlled $\r_n$ gate (see \cref{fig:4QFT})
	\begin{equation}\label{eq:Rn}
	\r_n = \r_\z(2\pi/2^n) =	\begin{pmatrix}
	1 & 0\\
	0 & e^{2\pi i/2^n}
	\end{pmatrix}.
	\end{equation}
	Such a gate can be implemented following the prescription in Ref. \cite{Kim2018}, in which different ways of decomposing the controlled $\r_n$ gate are derived, but we use the approach with one auxiliary qubit, such that three of the four diamond qubits are in use like so:
	\begin{equation}\label{eq:Rn_decomposed}
	\raisebox{1.7em}{\Qcircuit @C=0.3cm @R=0.3em @!R {& \ctrl{1}   & \qw\\
			& \gate{\r_n} & \qw\\
			\lstick{\ket{0}} & \qw      &  \rstick{\ket{0}} \qw}} 
	 \hspace{.7cm} = \hspace{.7cm}
	\raisebox{1.7em}{\Qcircuit @C=0.3cm @R=0.3em @!R { & \ctrl{1} & \qw & \ctrl{1} & \qw \\
	& \qswap & \qw & \qswap & \qw \\
	\lstick{\ket{0}} & \qswap\qwx & \gate{\r_n} & \qswap\qwx & \rstick{\ket{0}}\qw }}
	\end{equation}
	which locks one of the qubits in the $\ket{0}$ state. Our approach uses diamond instead of controlled-\cswap gates. 
	This introduces one extra qubit and, due to the symmetry of the diamond gate, we can pick this to be any of the four diamond qubits. We choose $C_2$ for this task and fix it in a $\ket{1}$ state. We then pick the other control qubit, $C_1$, as the control of the above $\r_n$gate, while we choose $T_1$ as the target and, finally, $T_2$ will be the ancillary qubit in the $\ket{0}$ state.  
	The circuit diagram takes the form
	\begin{equation}\label{eq:Rn_diamond}
	\hspace{0.75cm}\raisebox{2.5em}{\Qcircuit @C=0.3cm @R=0.3em @!R {
		\lstick{\ket{C_1}}& \ctrl{2}   & \qw\\
		\lstick{\ket{1}}& \qw & \qw \\
		\lstick{\ket{T_1}}& \gate{\r_n} & \qw\\
		\lstick{\ket{0}} & \qw      & \qw
	} }
	\hspace{.1cm} = \hspace{.9cm}
	\raisebox{2.5em}{\Qcircuit @C=0.3cm @R=0.3em @!R {
		\lstick{\ket{C_1}} & \multigate{3}{U(t_g)} & \qw & \multigate{3}{U(t_g)} & \qw \\
		\lstick{\ket{1}} & \ghost{U(t_g)} & \qw & \ghost{U(t_g)} & \qw \\
		\lstick{\ket{T_1}} & \ghost{U(t_g)} & \qw & \ghost{U(t_g)} & \qw \\
		\lstick{\ket{0}} & \ghost{U(t_g)} & \gate{\r_n} & \ghost{U(t_g)} &\qw
	}}
	\end{equation}
	Note that per \cref{eq:UTg}, the diamond gate introduces both a \cswap and a \cz gate, contrary to \cref{eq:Rn_decomposed}, where we only need a \cswap gate. However, the second diamond gate cancels out the \cz gate of the first diamond gate, making \cref{eq:Rn_diamond} equivalent to \cref{eq:Rn_decomposed}.
	
	In order to understand the circuit diagram in \cref{eq:Rn_diamond}, we consider the four different possibilities for two undetermined qubits $C_1$ and $T_1$, i.e., we consider the four different possibilities for the state $\ket{C_1T_1}$. Note that the four possibilities we consider are a complete basis for the two qubits. We want the qubit to obtain an overall factor $e^{2\pi i/2^n}$ only when $\ket{C_1T_1} = \ket{11}$. The four different cases are presented in \cref{fig:URUflowChart}, from which we see that the phase is only obtained in the desired case.
	
	We start with the one control and one target qubit in the state $\ket{C_2T_2} = \ket{10}$, while the two other qubits are in a arbitrary state $\ket{C_1T_1}$. In \cref{fig:URUflowChart}, we branch out to the four different possibilities $\ket{0100}$, $\ket{0110}$, $\ket{1100}$, and $\ket{1110}$. When we apply the first $U(t_g)$ gate, we effectively compare the states of control qubits $C_1$ and $C_2$ and if they are equal, the states of $T_1$ and $T_2$ are swapped. This happens in the fourth branch of \cref{fig:URUflowChart}. In the third branch, the states of the target qubits are the same, as is also the case for the control qubit. The states of the two target qubits do swap, and the same is the case for the two control qubits; however, as the states are the same for each pair of qubits, we do not observe this swapping. Due to the symmetry of the diamond gate, the target qubits' states are also compared and the states of the control qubits are swapped accordingly. This is what happens in the first branch. The target qubit states are different in the second branch, and no control state is swapped.
	When we apply the $R_n^{T_2}$ gate, only the fourth branch picks up the phase, as it is the only case for which $T_2$ is in the $\ket{1}$ state.
	Finally, when we apply the $U(t_g)$ gate again, we swap the states back and the qubits obtain an overall phase if and only if the original state is $\ket{1110}$.
	
	\subsection{Phase programming}
	
	Since the diamond gate is symmetric in the target and control qubits, we can place phase gates on all qubits between the two diamond gates and obtain controlled-phase gates on all qubits. Doing this, we obtain the following circuit:
	\begin{widetext}
		\begin{equation*}
			\hspace{0.5cm}\raisebox{3em}{\Qcircuit @C=0.3cm @R=0.3em @!R {
				\lstick{C_1}& \multigate{3}{U(t_g)} & \gate{\r_\z(t_1)} & \multigate{3}{U(t_g)} & \qw \\
				\lstick{C_2}& \ghost{U(t_G)} & \gate{\r_\z(t_2)} & \ghost{U(t_G)} & \qw \\
				\lstick{T_1}& \ghost{U(t_G)} & \gate{\r_\z(t_3)} & \ghost{U(t_G)} & \qw \\
				\lstick{T_2} & \ghost{U(t_G)} & \gate{\r_\z(t_4)} & \ghost{U(t_G)} & \qw
			}}
		 = 
		\raisebox{3em}{\Qcircuit @C=0.3cm @R=0.3em @!R {
			 & \gate{\r_\z(t_2)} & \gate{\r_\z(t_1)} & \gate{\r_\z(t_1)} & \gate{\r_\z(t_2)} & \ctrlo{1} & \ctrl{1} & \ctrlo{1} & \ctrl{1} & \qw \\
			 & \gate{\r_\z(t_1)}\qwx & \gate{\r_\z(t_2)}\qwx  & \gate{\r_\z(t_2)}\qwx & \gate{\r_\z(t_1)}\qwx & \ctrlo{1} & \ctrlo{1} & \ctrl{1} & \ctrl{1} & \qw \\
			 & \ctrlo{-1} & \ctrl{-1} & \ctrlo{-1} & \ctrl{-1} & \gate{\r_\z(t_4)} & \gate{\r_\z(t_3)} & \gate{\r_\z(t_3)} & \gate{\r_\z(t_4)} & \qw \\
			 & \ctrlo{-1} & \ctrlo{-1} & \ctrl{-1} & \ctrl{-1} & \gate{\r_\z(t_3)}\qwx & \gate{\r_\z(t_4)}\qwx & \gate{\r_\z(t_4)}\qwx & \gate{\r_\z(t_3)}\qwx & \qw
		}}
		\end{equation*}
	\end{widetext}
	This allows us to program what controlled-phase gates to keep, by fixing the states of some of the qubits, as well as choosing different values of $t_i$. If we set $t_1 = t_2 = t_3 = 0$ and $t_4 = 2\pi/2^n$ and initiate qubits $C_2$ and $T_2$ to $\ket{C_2T_2} = \ket{10}$, we obtain the result in \cref{eq:Rn_diamond} from above.
	Note that the open circles in the above circuit represent control qubits conditional on the control qubit being in the state $\ket{0}$. These are related to the solid circles, which are control qubits conditional on the qubit being the state $\ket{1}$, by performing a \x gate before and after the control qubit \cite{Nielsen2010}.

	\section{Native QFT algorithm with the diamond gate}\label{sec:QFT}
	
\begin{figure}
		\centering
		\includegraphics{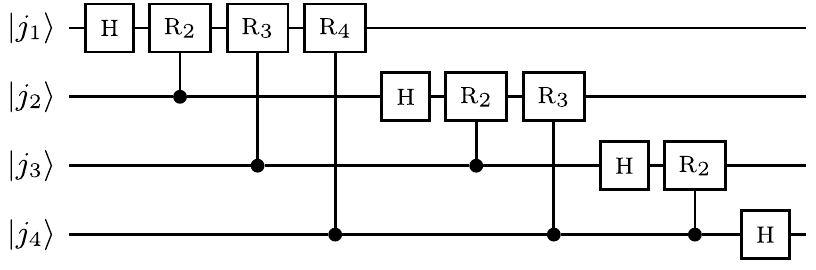}
	\caption{Quantum Fourier-transform circuit with four qubits \cite{Nielsen2010}.}
	\label{fig:4QFT}
	\end{figure}

	Having found a way to apply a controlled $\r_n$ gate to all qubits using the diamond gate, we are now ready to consider the problem of performing the quantum Fourier-transformation algorithm. This can be done using the circuit seen in \cref{fig:4QFT}. If $\ket{j}$ is a ket in an orthonormal basis, $\{\ket{0}, \ket{1}, \dots,\ket{N-1}\}$, this circuit performs the transformation
	\begin{equation}
		\ket{j} \mapsto \frac{1}{\sqrt{N}} \sum_{k=0}^{N-1} e^{2\pi ijk/2^n} \ket{k},
	\end{equation} 
	where $N = 2^n$, in which $n$ is the number of qubits, and $\ket{k}$ denotes the $k$th basis state \cite{Nielsen2010}.
	
\begin{figure}
		\centering
		\includegraphics[scale=0.7]{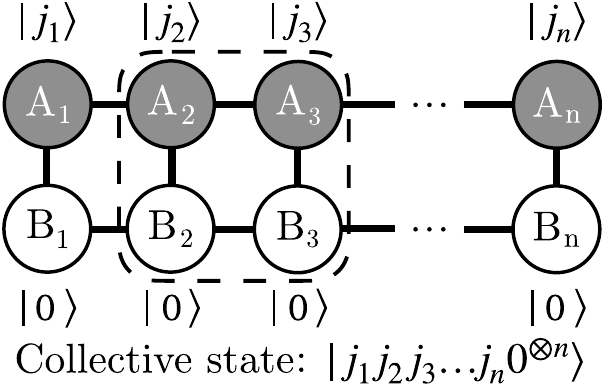}
		\caption{Two parallel interconnected lines of qubits are used in the QFT algorithm. The upper qubits (A) are initialized in the states $\ket{j_1}$,\dots, $\ket{j_n}$, while the lower qubits (B) are all in the $\ket{0}$ state. A diamond gate can be performed using one pair of A qubits and one pair of B qubits.}
		\label{fig:qubit-line}
	\end{figure}

	We must first arrange the qubits as we wish in order to perform the QFT algorithm using the diamond gate. As a starting point, we choose to arrange the qubits in two parallel lines as shown in \cref{fig:qubit-line}. The upper line contains the target qubits in arbitrary states $\ket{j_1},\dots,\ket{j_n}$, while the lower line shows the auxiliary qubits, all in states $\ket{0}$. Tuning a square of these qubits into resonance makes it possible to perform the diamond gate.
	This arrangement of qubits is a simple implementation of QFT; however, for many qubits, frequency crowding might lead to an unwanted crosstalk between qubits. In \cref{sec:altTopo}, we present alternative arrangements of qubits, which make this problem less severe.

	\subsection{QFT algorithm}
	
	Here, we present an algorithm for QFT using the topology presented in \cref{fig:qubit-line}:
	
	{\centering
		\begin{minipage}{.95\linewidth}
			
			\begin{algorithm}[H]
				\caption{Native-Gate-Assisted QFT on Double String of Qubits}
				\label{alg:dQFT}
				\begin{algorithmic}[1]
					\algnewcommand\algorithmicto{\textbf{to}}
					\algrenewtext{For}[3]%
					{\algorithmicfor\ $#1 = #2$ \algorithmicto\ $#3$ \algorithmicdo}
					\State Initialize qubits A$_1$,\dots,A$_n$ with the states to transform
					\State Initialize qubits B$_1$,\dots,B$_n$ in state $\ket{0}^{\otimes n}$
					\For{i}{1}{n}\Comment{For every target state}
					\State H on A$_i$
					\For{j}{i+1}{n}\Comment{\dots for every control state}
					\State X on B$_{j-1}$\Comment{Do $\Cr_n$ using native gate}
					\State $U(t_g)$ on (A$_j$, B$_{j-1}$, A$_{j-1}$, B$_j$)
					\State $\r_{j-i+1}$ on B$_j$
					\State $U(t_g)$ on (A$_j$, B$_{j-1}$, A$_{j-1}$, B$_j$)
					\State X on B$_{j-1}$
					\If{$j\neq n$}\Comment{Swap along chain (except last)}
					\State \ciswap{}($t_g/2$) on (A$_{j-1}$, A$_j$)
					\EndIf
					\EndFor
					\For{j}{n-1}{i+1}\Comment{Swap state back home}
					\State \ciswap{}($3t_g/2$) on (A$_{j-1}$, A$_j$)
					\EndFor
					\Comment{Done for A$_i$}
					\EndFor
					\State \text{Reverse order of A qubits.}
				\end{algorithmic}
			\end{algorithm}
		\end{minipage}
		\par
	}
	\bigskip
	
	The algorithm iterates through every A qubit, from A$_1$ (of state $\ket{j_1}$) to A$_n$ (of state $\ket{j_n}$). First, a Hadamard gate is applied to the A$_i$th qubit.
	Then a phase gate is applied to the same qubit, controlled by its neighboring A qubit (A$_{i+1}$), based upon the diamond-gate-assisted controlled $\r_n$ gate of \cref{eq:Rn_diamond}.
	This diamond gate is formed by tuning the diamond of A qubits (A$_i$ and A$_{i+1}$) and the associated pair of B qubits (B$_i$ and B$_{i+1}$).
	Then, the pair of A qubits swaps states through an \ciswap gate, such that another controlled phase may be applied to the A$_{i+1}$th qubit, which now holds the state of the A$_i$th qubit. This is repeated such that the original state of the A$_i$th qubit is swapped to the end of the chain. At this point, it is swapped back into place by doing the \ciswap{}s in reverse.
	
	Note that the initial swaps down the chain are \ciswap$(t_g/2)$, whereas the ones on the way back are \ciswap$(3t_g/2)$.
	Simply doing \ciswap$(t_g/2)$ twice could leave the pair of states with unwanted phases but using \ciswap$(3t_g/2)$ the second time around gets rid of these.
	Indeed, the trick here is that by swapping the states back through the chain, we ensure that each swapped pair of states will see each other twice.
	In turn, this will cancel any phase introduced by the initial \ciswap$(t_g/2)$.
	An example of this algorithm for $n=4$ can be found in \cref{sec:QFTalgo}.
	
	This QFT implementation on $O(n)$ input states requires $2n$ qubits.
	Thus, the number of qubits required grows linearly with the input size. The circuit depth grows as $O(n^2)$ as the standard QFT algorithm.
	On top of that, we can adapt the method of the approximate QFT and leave out the smallest of the $\r_n$ angles.
	This reduces the depth to grow as $O(n\log{n})$.
	Nevertheless, in absolute gate numbers, it does use more than regular QFT because we have to swap the states along the chain. The absolute gate counts are thus
	\begin{itemize}
		\setlength\itemsep{-.2em}
		\item \makebox[2cm][l]{$n$} \h gates
		\item \makebox[2cm][l]{$n(n+1)$} \x gates
		\item \makebox[2cm][l]{$n(n+1)/2$} $\r_n$ gates
		\item \makebox[2cm][l]{$n(n-1)$} \ciswap{} gates
	\end{itemize}
	This is without considering the QFT cross-swap of the A qubits at the end. Still, the algorithmic part of this procedure does not add asymptotic overhead to the existing QFT algorithm. In other words, the QFT algorithm with the diamond algorithm has neither an advantage nor a disadvantage over the QFT algorithm with standard gates; it is merely an alternative formulation of the algorithm.
	
	\subsection{Native-CNS-gate-based QFT algorithm}\label{sec:CNS_QFT}
	
	We wish to implement the QFT algorithm using the diamond version of the \cns gate in \cref{sec:cnsEquivalent}.
	
	Using an identity from Ref. \cite{Kim2018}, we replace each controlled $\r_n$ gate and introduce a pair of swaps to create two \cns gates:
	\begin{equation*}
		\raisebox{1.4em}{\Qcircuit @C=0.3cm @R=1em @!R {
			 & \gate{\r_n} & \qw \\
			 & \ctrl{-1}   & \qw
		} }
		 = 
		\raisebox{1.4em}{\Qcircuit @C=0.4cm @R=0.3em @!R {
			 & \gate{\r_{n+1}} & \targ{}   & \qswap & \qw                  & \qswap  & \targ{}   & \qw \\
			 & \gate{\r_{n+1}} & \ctrl{-1} & \qswap\qwx & \gate{\r_{n+1}^\dag} & \qswap\qwx & \ctrl{-1} & \qw
		}}
	\end{equation*}
	Now, the diamond circuit can only execute \cns gates on opposing qubits, and for the QFT algorithm in \cref{fig:4QFT}, we have to operate between all qubits.
	However, the \cns gate can also be used to make a pseudoswap gate for this purpose.
	Thus, we use pairs of \cns gates to swap the first qubit into place and back again as follows:
	\begin{equation*}
		\raisebox{2.25em}{\Qcircuit @C=0.3cm @R=0.2em @!R {
			& \targ{} & \qswap  & \qw  & \qswap & \targ{} & \qw\\
			& \qw  & \qw\qwx &  \qw & \qw\qwx & \qw & \qw \\
			& \ctrl{-2} & \qswap{}\qwx & \gate{\r_n^\dag} & \qswap{}\qwx & \ctrl{-2} & \qw
		} }
		= 
		\raisebox{2.25em}{\Qcircuit @C=0.3cm @R=0.2em @!R {
			& \ctrl{1} & \qswap & \qw & \qw  & \qw  & \qw & \qw & \qswap & \ctrl{1} & \qw\\
			& \targ{}  & \qswap{}\qwx & \targ{} & \qswap & \qw & \qswap & \targ{} & \qswap{}\qwx & \targ & \qw \\
			& \qw & \qw & \ctrl{-1} & \qswap{}\qwx & \gate{\r_n^\dag} & \qswap{}\qwx & \ctrl{-1} & \qw & \qw  & \qw
		}}
	\end{equation*}
	This works because of the symmetry of the swapping \cns gate pairs: the qubit at the control will only pick up a phase between each pair, and the target one is not transformed at all.
	Therefore, the effects of the \cns gates will cancel each other; however, we can still utilize the swapping functionality in between.
	
\begin{figure}
	\centering
	\includegraphics[scale=0.7]{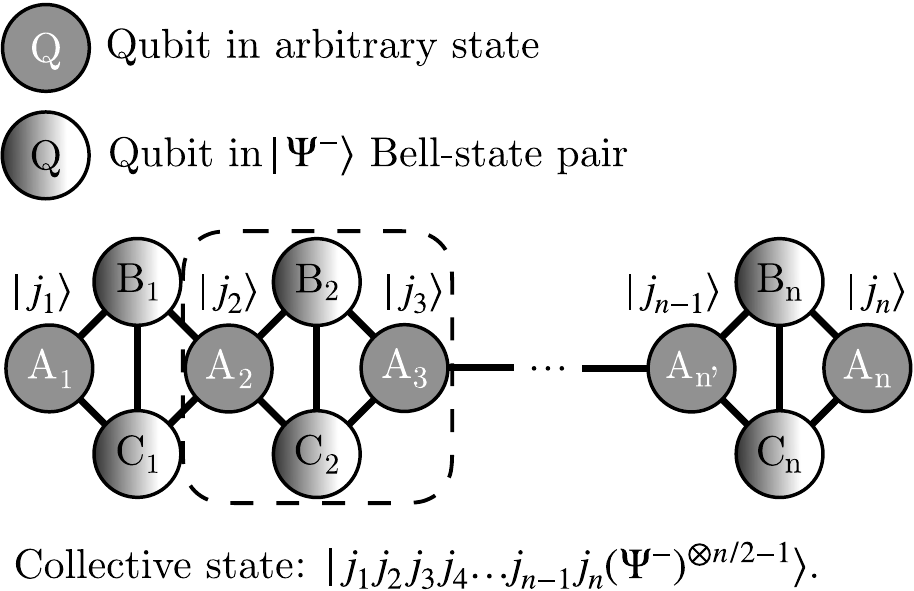}
	\caption{A chain of concatenated diamond circuits, each sharing a target qubit with its neighbor. This arrangement can be used to execute a \cns gate-assisted QFT algorithm.}
	\label{fig:qubit-concat}
	\end{figure}

	This algorithm can be run on a chain of diamond circuits, having the \cns gates only work on adjacent qubits.
	However, this sequence of gates can still be significantly refactored using the identity
	\begin{equation*}
		\raisebox{1.05em}{\Qcircuit @C=0.4cm @R=0.3em @!R {
			& \qswap & \targ{}   & \gate{\r_n} & \ctrl{1} & \qswap & \qw\\
			& \qswap\qwx & \ctrl{-1} & \qw         & \targ{}  & \qswap{}\qwx & \qw
		} }
		= 
		\raisebox{1.05em}{\Qcircuit @C=0.4cm @R=0.3em @!R {
				& \qswap & \ctrl{1} & \qw & \qw \\
				& \qswap\qwx & \targ{}  & \gate{\r_n} & \qw
		}}
	\end{equation*}
	Finally, this means that we can write the first part of the QFT algorithm in \cref{fig:4QFT}, which is the QFT of the first qubit, as
	\begin{widetext}
		\begin{equation*}
			\hspace{0.5cm}\raisebox{3.5em}{\Qcircuit @C=0.3cm @R=0.9em @!R {
			\lstick{\ket{j_1}} & \gate{\h} & \gate{\r_2} & \gate{\r_3} & \gate{\r_4} & \qw \\ 
			\lstick{\ket{j_2}} & \qw & \ctrl{-1} & \qw & \qw & \qw \\
			\lstick{\ket{j_3}} & \qw & \qw & \ctrl{-2} & \qw & \qw \\
			\lstick{\ket{j_4}} & \qw & \qw & \qw & \ctrl{-3} & \qw}}
			= \hspace{0.8cm}
			\raisebox{3.5em}{\Qcircuit @C=0.3cm @R=0.3em @!R {
				\lstick{\ket{j_1}} & \gate{\h} & \gate{\r_{3}} & \gate{\r_{4}} & \gate{\r_{5}} & \ctrl{1} & \qswap & \qw & \qw & \qw & \qw & \gate{\r_3^\dag} & \qw & \qw & \qw & \qw & \qswap & \ctrl{1} & \qw\\
				\lstick{\ket{j_2}} & \qw & \qw & \gate{\r_3}  & \qw & \targ{}  & \qswap{}\qwx & \ctrl{1} & \qswap & \qw & \qw & \gate{\r_4^\dag} & \qw & \qw & \qswap & \ctrl{1} & \qswap\qwx & \targ{} & \qw\\
				\lstick{\ket{j_3}} & \qw & \qw & \gate{\r_4} & \qw & \qw & \qw & \targ{} & \qswap{}\qwx & \targ{} & \qswap & \qw & \qswap & \targ{} & \qswap{}\qwx & \targ{}  & \qw & \qw      & \qw\\
				\lstick{\ket{j_4}} & \qw & \qw & \gate{\r_5} & \qw & \qw & \qw & \qw & \qw & \ctrl{-1} & \qswap{}\qwx & \gate{\r_5^\dag} & \qswap{}\qwx & \ctrl{-1} & \qw & \qw & \qw & \qw & \qw
			}}
		\end{equation*}
	\end{widetext}
	Now, three similar circuits have to be added to represent the QFT of all four qubits. In the same way, this method can be extended to more qubits.
	In particular, one hardware setup to execute this algorithm is seen in \cref{fig:qubit-concat},	which is just a sequence of concatenated diamond circuits, each sharing one qubit with its neighbor.
	Using this configuration, an additional QFT qubit requires three extra qubits in the circuit, i.e., QFT on $n$ qubits requires $3n+1$ circuit qubits.

\section{Quantum machine learning with the diamond gate}\label{eq:MLwDiamond}
	
	Another application where the diamond gate could prove helpful is in hybrid quantum-classical algorithms. Examples of this have already been shown with variational quantum eigensolvers \cite{Rasmussen2020c}. Here, we wish to consider the diamond gate in an HQC algorithm for machine learning on near-term quantum processors, namely quantum circuit learning as proposed by Ref. \cite{Mitarai2018}. Besides quantum circuit learning, we also wish to consider the problem of classifying data using PQCs with the diamond gate as an entangling gate. We do this following the approach taken by Ref. \cite{Hubregtsen2020}.
	
\subsection{Quantum circuit learning}\label{sec:quantumCircuitLearning}
	
\begin{figure*}
	\centering
	\includegraphics{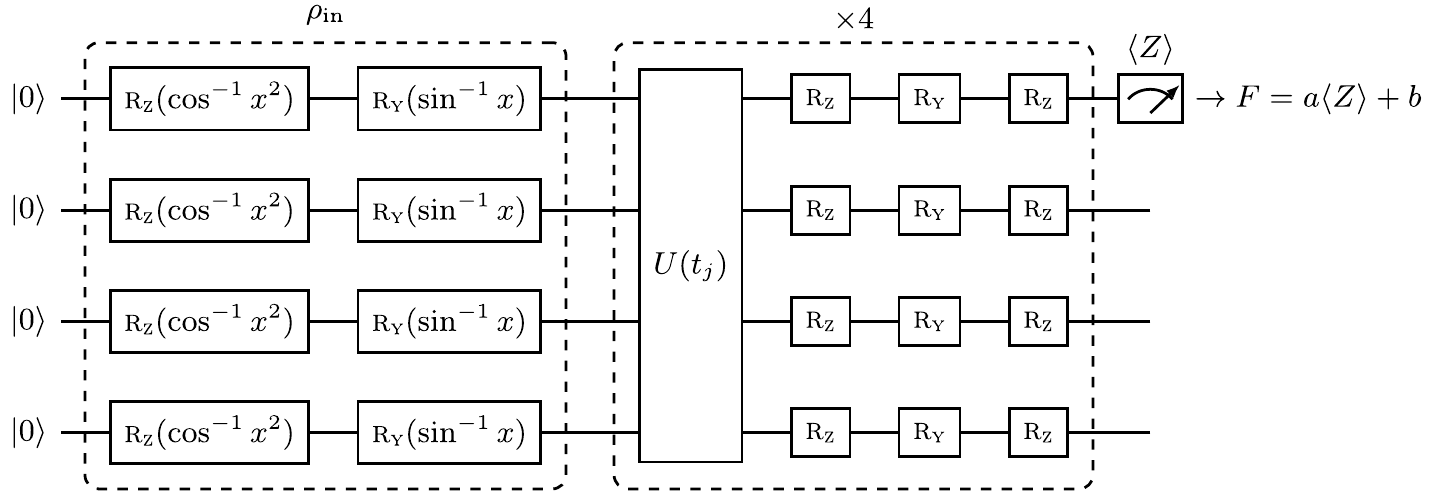}
	\caption{The circuit diagram of diamond-gate-aided circuit learning. The input data are encoded into the density matrix $\rho_\text{in}$ using $\r_\z$ and $\r_\y$ rotations on each qubit, which yield a state similar to \cref{eq:rho_in}. This happens in the left dashed box.
	The diamond gate, $U(t)$, acts as an entangling gate between the single-qubit rotations.
	The diamond-gate and single-qubit rotations in the right dashed box are repeated four times, with different rotations each time, yielding a four-layered parameterized quantum circuit.
	In the end, the first qubit is measured in the $Z$ basis and the output is scaled and shifted.
	We optimize the parameters $a$, $b$, and $t_j$, as well as the rotation angles of the single-qubit rotations in the right box.
	}\label{fig:1d_circuit}
\end{figure*}
	
	Quantum circuit learning is quantum-aided machine learning and can approximate nonlinear functions.
	In Ref. \cite{Mitarai2018}, it is argued that quantum circuit learning might have an advantage over classical approximators when it comes to more complex functions. This is because the computational cost of classical learning increases with the number of basis functions needed to represent complex functions. Quantum circuit learning, on the other hand, directly utilizes the exponential number of functions concerning the number of qubits to model the teacher. However, the approximation of such complex functions is beyond the capabilities of the current state-of-the-art quantum technology and we, therefore, approximate more straightforward functions.
	Quantum circuit learning is done in five steps: 
	\begin{enumerate}
	\item Encode input data in some quantum state $\ket{\psi_\text{in}(x)} = U_1(x)\ket{0\dots 0}$. 
	\item Apply parameterized quantum circuit to the input state $\ket{\psi_\text{out}(\bm\theta_i, x)} = U_2(\bm\theta_i)\ket{\psi_\text{in}(x)}$.
	\item Measure expectation values of some set of observables, $\{B_k\}$, by using some output function $F$, the output is defined to be $y(x, \bm\theta_i) = F(\{B_k(x, \bm \theta_i)\})$.	
	\item Minimize the cost function $L(f(x), y(x, \bm \theta_i))$ of the objective function $f(x)$ and output by tuning the parameters $\bm\theta_i$.
	\item Evaluate the performance by checking the value of the cost function or a similar evaluation criterion.
	\end{enumerate}
	
	We wish to use the diamond gate for approximating a function of one parameter $f: [-1, 1] \rightarrow \mathbb{R}$. We, therefore, encode a quantum state as the density matrix
	\begin{equation}\label{eq:rho_in}
		\rho_\text{in}(x) = \frac{1}{2^N} \bigotimes_{i=1}^N \left[I + xX_i + \sqrt{1-x^2}Z_i\right],
	\end{equation}
	where $X_i$ and $Z_i$ are X and Z rotations on the $i$th qubit, respectively. The density matrix can be generated by single-qubit rotations, $R_\y(\sin^{-1}x)$ and $R_\z(\cos^{-1}x^2)$ on each of the qubits. Our initial state is thus 
	\begin{equation}\label{eq:encoding1d}
	\ket{\psi_\text{in}(x)} = \prod_{i=1}^{N} \r_\y^{(i)}(\sin^{-1}x)\r^{(i)}_\z(\cos^{-1}x^2)\ket{0}^{\otimes N}.
	\end{equation}
	Note that in a practical numerical implementation, the continuous variable $x$ must be made discrete.
	This initialization has the effect that the tensor-product state will be a polynomial of order $N$ in $x$. However, the $x^N$ term is hidden with the observable $X^{\otimes N}$. So in order to access its factor, we must use an entangling gate, which in this case has to be the diamond gate, $U(t)$. In order to create a parameterized quantum circuit, we apply a general single-qubit rotation $U_\text{rot}(\bm\theta) = R_\z(\theta_1)R_\y(\theta_2)R_\z(\theta_3)$ to each of the qubits. The pair of gates, $U(t)$ and $U_\text{rot}(\bm\theta)$, are applied several times in layers, each with a different set of parameters. We use four such layers in our parameterized quantum circuit. At the end, we perform a measurement of the expectation value of $Z$ on the first qubit, which we scale and shift to produce the output function, $F(x, \bm\theta_i) = a\langle Z(x, \bm\theta_i) \rangle + b$, where $\theta_i$ is the $i$th tunable parameter of the circuit, from both $U(t)$ and $U_\text{rot}(\bm\theta)$. The resulting gate diagram can be seen in \cref{fig:1d_circuit}.
	
\begin{figure}
		\centering
		\includegraphics[width=\linewidth]{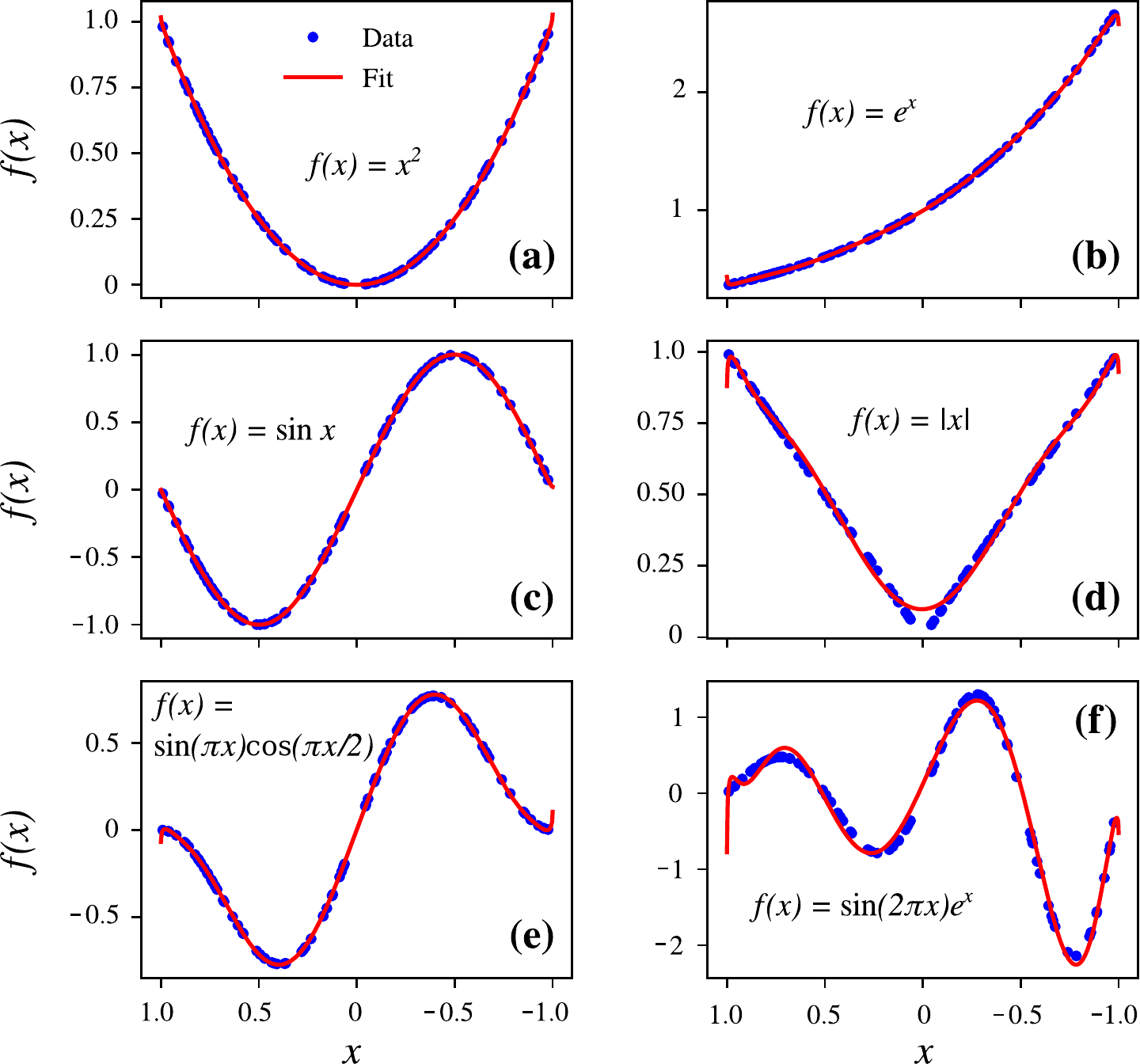}
		\caption{The data (blue points) and fit (red lines) obtained from the diamond-assisted quantum circuit learning using the quantum circuit seen in \cref{fig:1d_circuit}. The fit is given by the function $F(x,\bm \theta_i)$ described in the main text. The parameters, $\bm \theta_i$, are tuned to minimize the cost function for all $x$.
		The function, $f(x)$, used to sample the blue data points shown, is as follows: \textbf{(a)}
		$x^2$, \textbf{(b)} $e^x$, \textbf{(c)} $\sin{x}$, \textbf{(d)} $|x|$, \textbf{(e)} $\sin(\pi x)\cos(\pi x/2)$, and \textbf{(f)} $\sin(2\pi x) e^{x}$.}
		\label{fig:1d_results}
	\end{figure}

	We use the quadratic cost function, $L = ||f(x) - F(x, \bm\theta_i)||^2$ and optimize it using the \textsc{Adam} optimizer \cite{Kingma2014}. We test our quantum circuit using the same objective functions as in Ref. \cite{Mitarai2018}, namely $x^2$, $e^x$, $\sin x$, and $|x|$, as well as two more functions, $\sin(\pi x) \cos(\pi x/2)$ and $\sin (2\pi x)e^x$. We sample each objective function 100 times as training data. The results of the optimizations can be seen in \cref{fig:1d_results}. The parameterized quantum circuit approximates the function reasonably well, even though we only use four qubits and four layers, whereas in Ref. \cite{Mitarai2018}, six qubits and six layers are used to obtain the same precision. In Ref. \cite{Mitarai2018}, a fully connected transverse Ising model is used as the entangling gate, which is essentially equivalent to connecting all qubits with \ciswap gates, whereas we use the diamond gate to obtain the same results, showing that not only do we use fewer qubits and layers to achieve the same results but we also use fewer entangling operations per layer.
	
	The reason why we obtain the same results as Ref. \cite{Mitarai2018} with a more straightforward setup could be because the diamond gate is better at accessing the entire Hilbert space of the qubits compared to the transverse Ising model used in Ref. \cite{Mitarai2018}. However, it should be mentioned that we optimize the operation time, $t_j$, of the diamond gate for each of the layers, whereas in Ref. \cite{Mitarai2018} a constant time of operation is chosen for their entangling operation. This gives an extra optimization parameter per layer. This may not seem as much, especially when Ref. \cite{Mitarai2018} has more rotation gates, but optimization of the parameters of the entangling operation could be a considerable advantage in quantum circuits.
	
\subsection{Classification of data}
	
	Another scheme for quantum machine learning that employs PQC is data-point classification. We employ a PQC with the diamond gate as the entangling gate and use it to classify two-dimensional data points. We follow the approach in Ref. \cite{Hubregtsen2020} and use the same two-dimensional shapes in order to compare results directly.
	
	The classification of data follows the same algorithm as quantum circuit learning (see \cref{sec:quantumCircuitLearning}). However, here we use a different encoding, following the approach of Ref. \cite{Hubregtsen2020}. This so-called \emph{minimal expressive embedding} consists of applying single-qubit rotations like the encoding used previously. Thus instead of the encoding in \cref{eq:encoding1d}, we use the following:
	\begin{equation}\label{eq:encoding2d}
	\ket{\psi_\text{in}(x)} = \prod_{i=1}^{N} \r_\z^{(i)}(\pi/4)\r^{(i)}_\y(\pi/4)\r^{(i)}_\x(x_{i\text{ mod }2})\ket{0}^{\otimes N},
	\end{equation}
	where we first apply a rotation about the $x$ axis, which depends on $x_0$ and $x_1$ in an alternating fashion. We then perform rotations of $\pi/4$ first around the $y$ axis and then the $z$ axis for all qubits. Following this input encoding, we apply the PQC consisting of the diamond gate, $U(t)$, and a set of general single-qubit rotations $U_\text{rot}(\bm\theta)$ on each qubit. We apply these last two operations to the state several times, one for each layer that we want in our model. In the end, we measure the expectation value of $Z$ on the first qubit. The result is scaled, shifted, and then minimized using a binary cross-entropy function. Altogether, we perform the same manipulations as in \cref{fig:1d_circuit} but with a different encoding, $\rho_\text{in}$.
	
\begin{figure}
		\centering
		\includegraphics[width=\columnwidth]{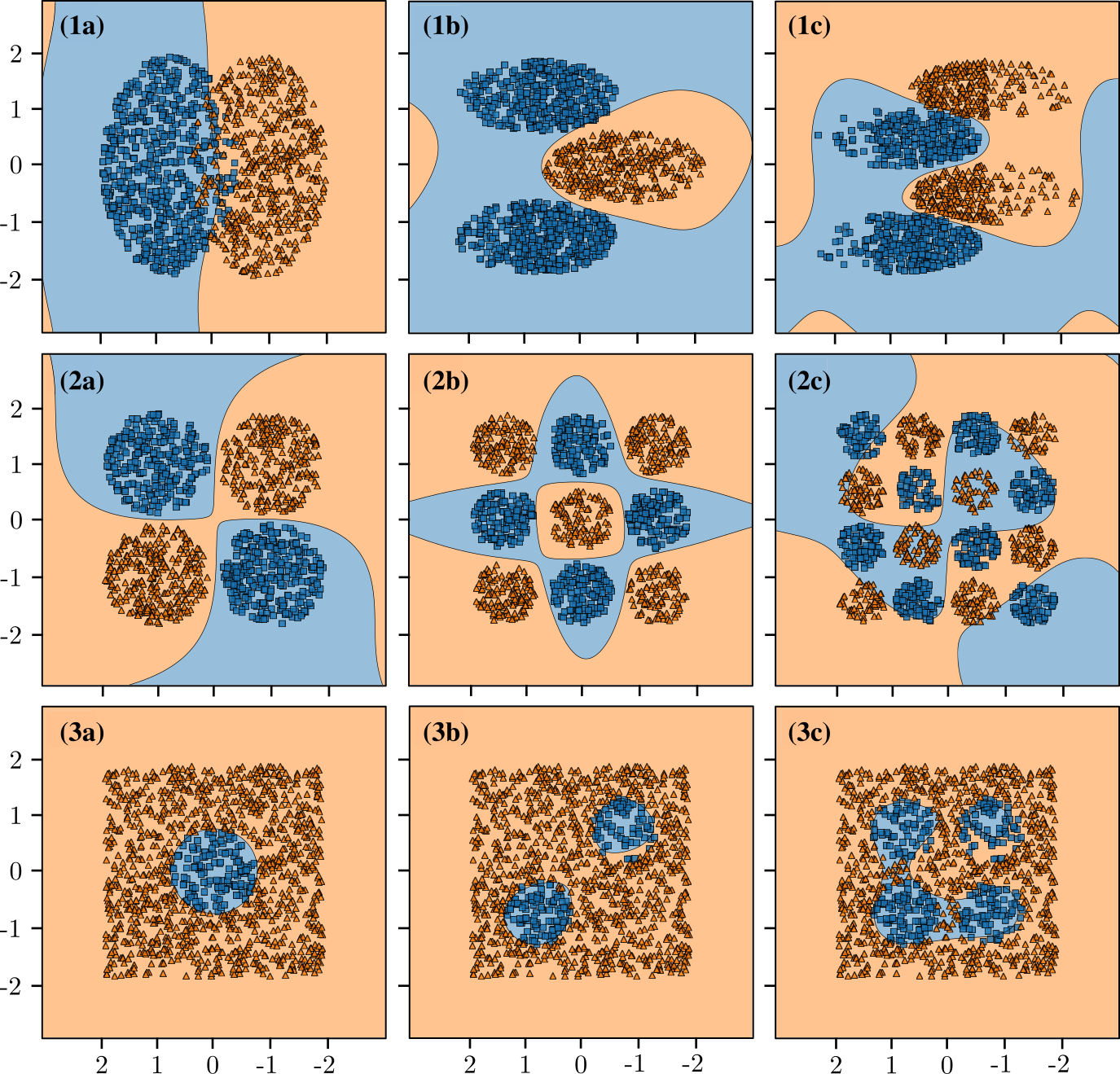}
		\caption{The diamond-assisted quantum circuit classification results of two-dimensional data points with two layers. The squares and triangles (blue and yellow) are the training data replicated from Ref. \cite{Hubregtsen2020}. The background colors indicate the predicted classification by the quantum circuit.}
		\label{fig:class_fits}
	\end{figure}

	In \cref{fig:class_fits}, we present the nine data sets as well as the classification results for two layers of $U(t)$ and $U_\text{rot}(\bm\theta)$. We find the data sets (1b), (2a), and (3a) to be perfectly classified for just a single layer, as well as for more layers. Data set (1a) achieves a classification accuracy just above 95\% for a single layer, which is the maximal for this data set due to the overlap of the data points. Data sets (2c) and (3b) also achieve a maximum classification accuracy for a single layer, of around 73\% and 98\%, respectively. Data sets (1c) and (2b) converge at a perfect accuracy for two layers, while the accuracy of the last data set (3c) converges for two layers to just above 90\%.
	
	We see almost perfect classification except for (2c) and (3c); however, these are most difficult due to their complexity. This performance is comparable to the most accurate two-layer networks used by Ref. \cite{Hubregtsen2020}. Surprisingly, the classification performance of (2c) and (3c) does not improve when the number of layers increases from two to three. This might be because the expressibility of the PQCs is already saturated at two layers, meaning that a further increase of the depth of the circuits does not increase the available Hilbert space of the circuit \cite{Sim2019,Rasmussen2021}.
	
\section{Conclusion and outlook}\label{sec:conclusion}
	
We show that the native diamond gate \cite{Loft2020} may be used beneficially compared to using standard gate sets, which makes it an auspicious architecture to realize on contemporary quantum technology platforms. 

First, we show that the diamond gate, together with single-qubit gates, can be turned into a controlled-\textsc{not} \cswap gate.
We then show how to symmetrically decompose the diamond gate into standard gates in two different ways.	
Also, by placing a diamond gate on either side of a phase gate, we create a controlled phase gate and argue that this could be used to obtain controlled-phase gates on all four qubits. Using this method, we present an algorithm for the quantum Fourier transform using only single-qubit gates and the diamond gate.

Finally, we apply the diamond gate in quantum machine-learning algorithms and show that it was helpful in a parameterized quantum circuit when used to approximate nonlinear functions and classify two-dimensional data.

We believe that these results outline the diamond gate as a possible building block in near-future quantum technologies. This, together with the fact that other quantum gates do not occur naturally in many quantum technology schemes, demonstrates that the applications of the diamond gate should be pursued further.

\begin{acknowledgements}
		We would like to thank L. B. Kristensen for discussions on different aspects of the work.
		This work is supported by the Danish Council for Independent Research and the Carlsberg Foundation, and is derived from E.B.'s Master's thesis \cite{Bahnsen2020}.
\end{acknowledgements}


\begin{thebibliography}{67}%
	\makeatletter
	\providecommand \@ifxundefined [1]{%
		\@ifx{#1\undefined}
	}%
	\providecommand \@ifnum [1]{%
		\ifnum #1\expandafter \@firstoftwo
		\else \expandafter \@secondoftwo
		\fi
	}%
	\providecommand \@ifx [1]{%
		\ifx #1\expandafter \@firstoftwo
		\else \expandafter \@secondoftwo
		\fi
	}%
	\providecommand \natexlab [1]{#1}%
	\providecommand \enquote  [1]{``#1''}%
	\providecommand \bibnamefont  [1]{#1}%
	\providecommand \bibfnamefont [1]{#1}%
	\providecommand \citenamefont [1]{#1}%
	\providecommand \href@noop [0]{\@secondoftwo}%
	\providecommand \href [0]{\begingroup \@sanitize@url \@href}%
	\providecommand \@href[1]{\@@startlink{#1}\@@href}%
	\providecommand \@@href[1]{\endgroup#1\@@endlink}%
	\providecommand \@sanitize@url [0]{\catcode `\\12\catcode `\$12\catcode
		`\&12\catcode `\#12\catcode `\^12\catcode `\_12\catcode `\%12\relax}%
	\providecommand \@@startlink[1]{}%
	\providecommand \@@endlink[0]{}%
	\providecommand \url  [0]{\begingroup\@sanitize@url \@url }%
	\providecommand \@url [1]{\endgroup\@href {#1}{\urlprefix }}%
	\providecommand \urlprefix  [0]{URL }%
	\providecommand \Eprint [0]{\href }%
	\providecommand \doibase [0]{http://dx.doi.org/}%
	\providecommand \selectlanguage [0]{\@gobble}%
	\providecommand \bibinfo  [0]{\@secondoftwo}%
	\providecommand \bibfield  [0]{\@secondoftwo}%
	\providecommand \translation [1]{[#1]}%
	\providecommand \BibitemOpen [0]{}%
	\providecommand \bibitemStop [0]{}%
	\providecommand \bibitemNoStop [0]{.\EOS\space}%
	\providecommand \EOS [0]{\spacefactor3000\relax}%
	\providecommand \BibitemShut  [1]{\csname bibitem#1\endcsname}%
	\let\auto@bib@innerbib\@empty
	\bibitem [{\citenamefont {Arute}\ \emph {et~al.}(2019)\citenamefont {Arute},
		\citenamefont {Arya}, \citenamefont {Babbush}, \citenamefont {Bacon},
		\citenamefont {Bardin}, \citenamefont {Barends}, \citenamefont {Biswas},
		\citenamefont {Boixo}, \citenamefont {Brandao}, \citenamefont {Buell},
		\citenamefont {Burkett}, \citenamefont {Chen}, \citenamefont {Chen},
		\citenamefont {Chiaro}, \citenamefont {Collins}, \citenamefont {Courtney},
		\citenamefont {Dunsworth}, \citenamefont {Farhi}, \citenamefont {Foxen},
		\citenamefont {Fowler}, \citenamefont {Gidney}, \citenamefont {Giustina},
		\citenamefont {Graff}, \citenamefont {Guerin}, \citenamefont {Habegger},
		\citenamefont {Harrigan}, \citenamefont {Hartmann}, \citenamefont {Ho},
		\citenamefont {Hoffmann}, \citenamefont {Huang}, \citenamefont {Humble},
		\citenamefont {Isakov}, \citenamefont {Jeffrey}, \citenamefont {Jiang},
		\citenamefont {Kafri}, \citenamefont {Kechedzhi}, \citenamefont {Kelly},
		\citenamefont {Klimov}, \citenamefont {Knysh}, \citenamefont {Korotkov},
		\citenamefont {Kostritsa}, \citenamefont {Landhuis}, \citenamefont
		{Lindmark}, \citenamefont {Lucero}, \citenamefont {Lyakh}, \citenamefont
		{Mandr{\`a}}, \citenamefont {McClean}, \citenamefont {McEwen}, \citenamefont
		{Megrant}, \citenamefont {Mi}, \citenamefont {Michielsen}, \citenamefont
		{Mohseni}, \citenamefont {Mutus}, \citenamefont {Naaman}, \citenamefont
		{Neeley}, \citenamefont {Neill}, \citenamefont {Niu}, \citenamefont {Ostby},
		\citenamefont {Petukhov}, \citenamefont {Platt}, \citenamefont {Quintana},
		\citenamefont {Rieffel}, \citenamefont {Roushan}, \citenamefont {Rubin},
		\citenamefont {Sank}, \citenamefont {Satzinger}, \citenamefont {Smelyanskiy},
		\citenamefont {Sung}, \citenamefont {Trevithick}, \citenamefont
		{Vainsencher}, \citenamefont {Villalonga}, \citenamefont {White},
		\citenamefont {Yao}, \citenamefont {Yeh}, \citenamefont {Zalcman},
		\citenamefont {Neven},\ and\ \citenamefont {Martinis}}]{Arute2019}%
	\BibitemOpen
	\bibfield  {author} {\bibinfo {author} {\bibfnamefont {F.}~\bibnamefont
			{Arute}}, \bibinfo {author} {\bibfnamefont {K.}~\bibnamefont {Arya}},
		\bibinfo {author} {\bibfnamefont {R.}~\bibnamefont {Babbush}}, \bibinfo
		{author} {\bibfnamefont {D.}~\bibnamefont {Bacon}}, \bibinfo {author}
		{\bibfnamefont {J.~C.}\ \bibnamefont {Bardin}}, \bibinfo {author}
		{\bibfnamefont {R.}~\bibnamefont {Barends}}, \bibinfo {author} {\bibfnamefont
			{R.}~\bibnamefont {Biswas}}, \bibinfo {author} {\bibfnamefont
			{S.}~\bibnamefont {Boixo}}, \bibinfo {author} {\bibfnamefont {F.~G. S.~L.}\
			\bibnamefont {Brandao}}, \bibinfo {author} {\bibfnamefont {D.~A.}\
			\bibnamefont {Buell}}, \bibinfo {author} {\bibfnamefont {B.}~\bibnamefont
			{Burkett}}, \bibinfo {author} {\bibfnamefont {Y.}~\bibnamefont {Chen}},
		\bibinfo {author} {\bibfnamefont {Z.}~\bibnamefont {Chen}}, \bibinfo {author}
		{\bibfnamefont {B.}~\bibnamefont {Chiaro}}, \bibinfo {author} {\bibfnamefont
			{R.}~\bibnamefont {Collins}}, \bibinfo {author} {\bibfnamefont
			{W.}~\bibnamefont {Courtney}}, \bibinfo {author} {\bibfnamefont
			{A.}~\bibnamefont {Dunsworth}}, \bibinfo {author} {\bibfnamefont
			{E.}~\bibnamefont {Farhi}}, \bibinfo {author} {\bibfnamefont
			{B.}~\bibnamefont {Foxen}}, \bibinfo {author} {\bibfnamefont
			{A.}~\bibnamefont {Fowler}}, \bibinfo {author} {\bibfnamefont
			{C.}~\bibnamefont {Gidney}}, \bibinfo {author} {\bibfnamefont
			{M.}~\bibnamefont {Giustina}}, \bibinfo {author} {\bibfnamefont
			{R.}~\bibnamefont {Graff}}, \bibinfo {author} {\bibfnamefont
			{K.}~\bibnamefont {Guerin}}, \bibinfo {author} {\bibfnamefont
			{S.}~\bibnamefont {Habegger}}, \bibinfo {author} {\bibfnamefont {M.~P.}\
			\bibnamefont {Harrigan}}, \bibinfo {author} {\bibfnamefont {M.~J.}\
			\bibnamefont {Hartmann}}, \bibinfo {author} {\bibfnamefont {A.}~\bibnamefont
			{Ho}}, \bibinfo {author} {\bibfnamefont {M.}~\bibnamefont {Hoffmann}},
		\bibinfo {author} {\bibfnamefont {T.}~\bibnamefont {Huang}}, \bibinfo
		{author} {\bibfnamefont {T.~S.}\ \bibnamefont {Humble}}, \bibinfo {author}
		{\bibfnamefont {S.~V.}\ \bibnamefont {Isakov}}, \bibinfo {author}
		{\bibfnamefont {E.}~\bibnamefont {Jeffrey}}, \bibinfo {author} {\bibfnamefont
			{Z.}~\bibnamefont {Jiang}}, \bibinfo {author} {\bibfnamefont
			{D.}~\bibnamefont {Kafri}}, \bibinfo {author} {\bibfnamefont
			{K.}~\bibnamefont {Kechedzhi}}, \bibinfo {author} {\bibfnamefont
			{J.}~\bibnamefont {Kelly}}, \bibinfo {author} {\bibfnamefont {P.~V.}\
			\bibnamefont {Klimov}}, \bibinfo {author} {\bibfnamefont {S.}~\bibnamefont
			{Knysh}}, \bibinfo {author} {\bibfnamefont {A.}~\bibnamefont {Korotkov}},
		\bibinfo {author} {\bibfnamefont {F.}~\bibnamefont {Kostritsa}}, \bibinfo
		{author} {\bibfnamefont {D.}~\bibnamefont {Landhuis}}, \bibinfo {author}
		{\bibfnamefont {M.}~\bibnamefont {Lindmark}}, \bibinfo {author}
		{\bibfnamefont {E.}~\bibnamefont {Lucero}}, \bibinfo {author} {\bibfnamefont
			{D.}~\bibnamefont {Lyakh}}, \bibinfo {author} {\bibfnamefont
			{S.}~\bibnamefont {Mandr{\`a}}}, \bibinfo {author} {\bibfnamefont {J.~R.}\
			\bibnamefont {McClean}}, \bibinfo {author} {\bibfnamefont {M.}~\bibnamefont
			{McEwen}}, \bibinfo {author} {\bibfnamefont {A.}~\bibnamefont {Megrant}},
		\bibinfo {author} {\bibfnamefont {X.}~\bibnamefont {Mi}}, \bibinfo {author}
		{\bibfnamefont {K.}~\bibnamefont {Michielsen}}, \bibinfo {author}
		{\bibfnamefont {M.}~\bibnamefont {Mohseni}}, \bibinfo {author} {\bibfnamefont
			{J.}~\bibnamefont {Mutus}}, \bibinfo {author} {\bibfnamefont
			{O.}~\bibnamefont {Naaman}}, \bibinfo {author} {\bibfnamefont
			{M.}~\bibnamefont {Neeley}}, \bibinfo {author} {\bibfnamefont
			{C.}~\bibnamefont {Neill}}, \bibinfo {author} {\bibfnamefont {M.~Y.}\
			\bibnamefont {Niu}}, \bibinfo {author} {\bibfnamefont {E.}~\bibnamefont
			{Ostby}}, \bibinfo {author} {\bibfnamefont {A.}~\bibnamefont {Petukhov}},
		\bibinfo {author} {\bibfnamefont {J.~C.}\ \bibnamefont {Platt}}, \bibinfo
		{author} {\bibfnamefont {C.}~\bibnamefont {Quintana}}, \bibinfo {author}
		{\bibfnamefont {E.~G.}\ \bibnamefont {Rieffel}}, \bibinfo {author}
		{\bibfnamefont {P.}~\bibnamefont {Roushan}}, \bibinfo {author} {\bibfnamefont
			{N.~C.}\ \bibnamefont {Rubin}}, \bibinfo {author} {\bibfnamefont
			{D.}~\bibnamefont {Sank}}, \bibinfo {author} {\bibfnamefont {K.~J.}\
			\bibnamefont {Satzinger}}, \bibinfo {author} {\bibfnamefont {V.}~\bibnamefont
			{Smelyanskiy}}, \bibinfo {author} {\bibfnamefont {K.~J.}\ \bibnamefont
			{Sung}}, \bibinfo {author} {\bibfnamefont {M.~D.}\ \bibnamefont
			{Trevithick}}, \bibinfo {author} {\bibfnamefont {A.}~\bibnamefont
			{Vainsencher}}, \bibinfo {author} {\bibfnamefont {B.}~\bibnamefont
			{Villalonga}}, \bibinfo {author} {\bibfnamefont {T.}~\bibnamefont {White}},
		\bibinfo {author} {\bibfnamefont {Z.~J.}\ \bibnamefont {Yao}}, \bibinfo
		{author} {\bibfnamefont {P.}~\bibnamefont {Yeh}}, \bibinfo {author}
		{\bibfnamefont {A.}~\bibnamefont {Zalcman}}, \bibinfo {author} {\bibfnamefont
			{H.}~\bibnamefont {Neven}}, \ and\ \bibinfo {author} {\bibfnamefont {J.~M.}\
			\bibnamefont {Martinis}},\ }\href {\doibase 10.1038/s41586-019-1666-5}
	{\bibfield  {journal} {\bibinfo  {journal} {Nature}\ }\textbf {\bibinfo
			{volume} {574}},\ \bibinfo {pages} {505} (\bibinfo {year}
		{2019})}\BibitemShut {NoStop}%
	\bibitem [{\citenamefont {Bravyi}\ \emph {et~al.}(2020)\citenamefont {Bravyi},
		\citenamefont {Gosset}, \citenamefont {K{\"o}nig},\ and\ \citenamefont
		{Tomamichel}}]{Bravyi2020}%
	\BibitemOpen
	\bibfield  {author} {\bibinfo {author} {\bibfnamefont {S.}~\bibnamefont
			{Bravyi}}, \bibinfo {author} {\bibfnamefont {D.}~\bibnamefont {Gosset}},
		\bibinfo {author} {\bibfnamefont {R.}~\bibnamefont {K{\"o}nig}}, \ and\
		\bibinfo {author} {\bibfnamefont {M.}~\bibnamefont {Tomamichel}},\ }\href
	{\doibase 10.1038/s41567-020-0948-z} {\bibfield  {journal} {\bibinfo
			{journal} {Nature Physics}\ } (\bibinfo {year} {2020}),\
		10.1038/s41567-020-0948-z}\BibitemShut {NoStop}%
	\bibitem [{\citenamefont {Preskill}(2018)}]{Preskill2018}%
	\BibitemOpen
	\bibfield  {author} {\bibinfo {author} {\bibfnamefont {J.}~\bibnamefont
			{Preskill}},\ }\href {\doibase 10.22331/q-2018-08-06-79} {\bibfield
		{journal} {\bibinfo  {journal} {{Quantum}}\ }\textbf {\bibinfo {volume}
			{2}},\ \bibinfo {pages} {79} (\bibinfo {year} {2018})}\BibitemShut {NoStop}%
	\bibitem [{\citenamefont {Loft}\ \emph {et~al.}(2020)\citenamefont {Loft},
		\citenamefont {Kjaergaard}, \citenamefont {Kristensen}, \citenamefont
		{Andersen}, \citenamefont {Larsen}, \citenamefont {Gustavsson}, \citenamefont
		{Oliver},\ and\ \citenamefont {Zinner}}]{Loft2020}%
	\BibitemOpen
	\bibfield  {author} {\bibinfo {author} {\bibfnamefont {N.~J.~S.}\
			\bibnamefont {Loft}}, \bibinfo {author} {\bibfnamefont {M.}~\bibnamefont
			{Kjaergaard}}, \bibinfo {author} {\bibfnamefont {L.~B.}\ \bibnamefont
			{Kristensen}}, \bibinfo {author} {\bibfnamefont {C.~K.}\ \bibnamefont
			{Andersen}}, \bibinfo {author} {\bibfnamefont {T.~W.}\ \bibnamefont
			{Larsen}}, \bibinfo {author} {\bibfnamefont {S.}~\bibnamefont {Gustavsson}},
		\bibinfo {author} {\bibfnamefont {W.~D.}\ \bibnamefont {Oliver}}, \ and\
		\bibinfo {author} {\bibfnamefont {N.~T.}\ \bibnamefont {Zinner}},\ }\href
	{\doibase 10.1038/s41534-020-0275-3} {\bibfield  {journal} {\bibinfo
			{journal} {npj Quantum Information}\ }\textbf {\bibinfo {volume} {6}},\
		\bibinfo {pages} {47} (\bibinfo {year} {2020})}\BibitemShut {NoStop}%
	\bibitem [{\citenamefont {Nielsen}\ and\ \citenamefont
		{Chuang}(2010)}]{Nielsen2010}%
	\BibitemOpen
	\bibfield  {author} {\bibinfo {author} {\bibfnamefont {M.~A.}\ \bibnamefont
			{Nielsen}}\ and\ \bibinfo {author} {\bibfnamefont {I.~L.}\ \bibnamefont
			{Chuang}},\ }\href@noop {} {\emph {\bibinfo {title} {Quantum Computation and
				Quantum Information}}}\ (\bibinfo  {publisher} {Cambridge University Press},\
	\bibinfo {address} {Cambridge, UK},\ \bibinfo {year} {2010})\BibitemShut
	{NoStop}%
	\bibitem [{\citenamefont {Milburn}(1989)}]{Milburn1989}%
	\BibitemOpen
	\bibfield  {author} {\bibinfo {author} {\bibfnamefont {G.~J.}\ \bibnamefont
			{Milburn}},\ }\href {\doibase 10.1103/PhysRevLett.62.2124} {\bibfield
		{journal} {\bibinfo  {journal} {Phys. Rev. Lett.}\ }\textbf {\bibinfo
			{volume} {62}},\ \bibinfo {pages} {2124} (\bibinfo {year}
		{1989})}\BibitemShut {NoStop}%
	\bibitem [{\citenamefont {Chau}\ and\ \citenamefont
		{Wilczek}(1995)}]{Chau1995}%
	\BibitemOpen
	\bibfield  {author} {\bibinfo {author} {\bibfnamefont {H.~F.}\ \bibnamefont
			{Chau}}\ and\ \bibinfo {author} {\bibfnamefont {F.}~\bibnamefont {Wilczek}},\
	}\href {\doibase 10.1103/PhysRevLett.75.748} {\bibfield  {journal} {\bibinfo
			{journal} {Phys. Rev. Lett.}\ }\textbf {\bibinfo {volume} {75}},\ \bibinfo
		{pages} {748} (\bibinfo {year} {1995})}\BibitemShut {NoStop}%
	\bibitem [{\citenamefont {Fiur\'a\ifmmode~\check{s}\else
			\v{s}\fi{}ek}(2006)}]{Fiuraifmmode2006}%
	\BibitemOpen
	\bibfield  {author} {\bibinfo {author} {\bibfnamefont {J.}~\bibnamefont
			{Fiur\'a\ifmmode~\check{s}\else \v{s}\fi{}ek}},\ }\href {\doibase
		10.1103/PhysRevA.73.062313} {\bibfield  {journal} {\bibinfo  {journal} {Phys.
				Rev. A}\ }\textbf {\bibinfo {volume} {73}},\ \bibinfo {pages} {062313}
		(\bibinfo {year} {2006})}\BibitemShut {NoStop}%
	\bibitem [{\citenamefont {Fiur\'a\ifmmode~\check{s}\else
			\v{s}\fi{}ek}(2008)}]{Fiuraifmmode2008}%
	\BibitemOpen
	\bibfield  {author} {\bibinfo {author} {\bibfnamefont {J.}~\bibnamefont
			{Fiur\'a\ifmmode~\check{s}\else \v{s}\fi{}ek}},\ }\href {\doibase
		10.1103/PhysRevA.78.032317} {\bibfield  {journal} {\bibinfo  {journal} {Phys.
				Rev. A}\ }\textbf {\bibinfo {volume} {78}},\ \bibinfo {pages} {032317}
		(\bibinfo {year} {2008})}\BibitemShut {NoStop}%
	\bibitem [{\citenamefont {Gong}\ \emph {et~al.}(2008)\citenamefont {Gong},
		\citenamefont {Guo},\ and\ \citenamefont {Ralph}}]{Gong2008}%
	\BibitemOpen
	\bibfield  {author} {\bibinfo {author} {\bibfnamefont {Y.-X.}\ \bibnamefont
			{Gong}}, \bibinfo {author} {\bibfnamefont {G.-C.}\ \bibnamefont {Guo}}, \
		and\ \bibinfo {author} {\bibfnamefont {T.~C.}\ \bibnamefont {Ralph}},\ }\href
	{\doibase 10.1103/PhysRevA.78.012305} {\bibfield  {journal} {\bibinfo
			{journal} {Phys. Rev. A}\ }\textbf {\bibinfo {volume} {78}},\ \bibinfo
		{pages} {012305} (\bibinfo {year} {2008})}\BibitemShut {NoStop}%
	\bibitem [{\citenamefont {Patel}\ \emph {et~al.}(2016)\citenamefont {Patel},
		\citenamefont {Ho}, \citenamefont {Ferreyrol}, \citenamefont {Ralph},\ and\
		\citenamefont {Pryde}}]{Patel2016}%
	\BibitemOpen
	\bibfield  {author} {\bibinfo {author} {\bibfnamefont {R.~B.}\ \bibnamefont
			{Patel}}, \bibinfo {author} {\bibfnamefont {J.}~\bibnamefont {Ho}}, \bibinfo
		{author} {\bibfnamefont {F.}~\bibnamefont {Ferreyrol}}, \bibinfo {author}
		{\bibfnamefont {T.~C.}\ \bibnamefont {Ralph}}, \ and\ \bibinfo {author}
		{\bibfnamefont {G.~J.}\ \bibnamefont {Pryde}},\ }\href {\doibase
		10.1126/sciadv.1501531} {\bibfield  {journal} {\bibinfo  {journal} {Science
				Advances}\ }\textbf {\bibinfo {volume} {2}} (\bibinfo {year} {2016}),\
		10.1126/sciadv.1501531}\BibitemShut {NoStop}%
	\bibitem [{\citenamefont {Ono}\ \emph {et~al.}(2017)\citenamefont {Ono},
		\citenamefont {Okamoto}, \citenamefont {Tanida}, \citenamefont {Hofmann},\
		and\ \citenamefont {Takeuchi}}]{Ono2017}%
	\BibitemOpen
	\bibfield  {author} {\bibinfo {author} {\bibfnamefont {T.}~\bibnamefont
			{Ono}}, \bibinfo {author} {\bibfnamefont {R.}~\bibnamefont {Okamoto}},
		\bibinfo {author} {\bibfnamefont {M.}~\bibnamefont {Tanida}}, \bibinfo
		{author} {\bibfnamefont {H.~F.}\ \bibnamefont {Hofmann}}, \ and\ \bibinfo
		{author} {\bibfnamefont {S.}~\bibnamefont {Takeuchi}},\ }\href
	{https://doi.org/10.1038/srep45353} {\bibfield  {journal} {\bibinfo
			{journal} {Scientific Reports}\ }\textbf {\bibinfo {volume} {7}},\ \bibinfo
		{pages} {45353 EP } (\bibinfo {year} {2017})},\ \bibinfo {note}
	{article}\BibitemShut {NoStop}%
	\bibitem [{\citenamefont {Smolin}\ and\ \citenamefont
		{DiVincenzo}(1996)}]{Smolin1996}%
	\BibitemOpen
	\bibfield  {author} {\bibinfo {author} {\bibfnamefont {J.~A.}\ \bibnamefont
			{Smolin}}\ and\ \bibinfo {author} {\bibfnamefont {D.~P.}\ \bibnamefont
			{DiVincenzo}},\ }\href {\doibase 10.1103/PhysRevA.53.2855} {\bibfield
		{journal} {\bibinfo  {journal} {Phys. Rev. A}\ }\textbf {\bibinfo {volume}
			{53}},\ \bibinfo {pages} {2855} (\bibinfo {year} {1996})}\BibitemShut
	{NoStop}%
	\bibitem [{\citenamefont {B{\ae}kkegaard}\ \emph {et~al.}(2019)\citenamefont
		{B{\ae}kkegaard}, \citenamefont {Kristensen}, \citenamefont {Loft},
		\citenamefont {Andersen}, \citenamefont {Petrosyan},\ and\ \citenamefont
		{Zinner}}]{Baekkegaard2019}%
	\BibitemOpen
	\bibfield  {author} {\bibinfo {author} {\bibfnamefont {T.}~\bibnamefont
			{B{\ae}kkegaard}}, \bibinfo {author} {\bibfnamefont {L.~B.}\ \bibnamefont
			{Kristensen}}, \bibinfo {author} {\bibfnamefont {N.~J.~S.}\ \bibnamefont
			{Loft}}, \bibinfo {author} {\bibfnamefont {C.~K.}\ \bibnamefont {Andersen}},
		\bibinfo {author} {\bibfnamefont {D.}~\bibnamefont {Petrosyan}}, \ and\
		\bibinfo {author} {\bibfnamefont {N.~T.}\ \bibnamefont {Zinner}},\ }\href
	{\doibase 10.1038/s41598-019-49657-1} {\bibfield  {journal} {\bibinfo
			{journal} {Scientific Reports}\ }\textbf {\bibinfo {volume} {9}},\ \bibinfo
		{pages} {13389} (\bibinfo {year} {2019})}\BibitemShut {NoStop}%
	\bibitem [{\citenamefont {Poletto}\ \emph {et~al.}(2012)\citenamefont
		{Poletto}, \citenamefont {Gambetta}, \citenamefont {Merkel}, \citenamefont
		{Smolin}, \citenamefont {Chow}, \citenamefont {C\'orcoles}, \citenamefont
		{Keefe}, \citenamefont {Rothwell}, \citenamefont {Rozen}, \citenamefont
		{Abraham}, \citenamefont {Rigetti},\ and\ \citenamefont
		{Steffen}}]{Poletto2012}%
	\BibitemOpen
	\bibfield  {author} {\bibinfo {author} {\bibfnamefont {S.}~\bibnamefont
			{Poletto}}, \bibinfo {author} {\bibfnamefont {J.~M.}\ \bibnamefont
			{Gambetta}}, \bibinfo {author} {\bibfnamefont {S.~T.}\ \bibnamefont
			{Merkel}}, \bibinfo {author} {\bibfnamefont {J.~A.}\ \bibnamefont {Smolin}},
		\bibinfo {author} {\bibfnamefont {J.~M.}\ \bibnamefont {Chow}}, \bibinfo
		{author} {\bibfnamefont {A.~D.}\ \bibnamefont {C\'orcoles}}, \bibinfo
		{author} {\bibfnamefont {G.~A.}\ \bibnamefont {Keefe}}, \bibinfo {author}
		{\bibfnamefont {M.~B.}\ \bibnamefont {Rothwell}}, \bibinfo {author}
		{\bibfnamefont {J.~R.}\ \bibnamefont {Rozen}}, \bibinfo {author}
		{\bibfnamefont {D.~W.}\ \bibnamefont {Abraham}}, \bibinfo {author}
		{\bibfnamefont {C.}~\bibnamefont {Rigetti}}, \ and\ \bibinfo {author}
		{\bibfnamefont {M.}~\bibnamefont {Steffen}},\ }\href {\doibase
		10.1103/PhysRevLett.109.240505} {\bibfield  {journal} {\bibinfo  {journal}
			{Phys. Rev. Lett.}\ }\textbf {\bibinfo {volume} {109}},\ \bibinfo {pages}
		{240505} (\bibinfo {year} {2012})}\BibitemShut {NoStop}%
	\bibitem [{\citenamefont {Rasmussen}\ \emph {et~al.}(2019)\citenamefont
		{Rasmussen}, \citenamefont {Christensen},\ and\ \citenamefont
		{Zinner}}]{Rasmussen2019}%
	\BibitemOpen
	\bibfield  {author} {\bibinfo {author} {\bibfnamefont {S.~E.}\ \bibnamefont
			{Rasmussen}}, \bibinfo {author} {\bibfnamefont {K.~S.}\ \bibnamefont
			{Christensen}}, \ and\ \bibinfo {author} {\bibfnamefont {N.~T.}\ \bibnamefont
			{Zinner}},\ }\href {\doibase 10.1103/PhysRevB.99.134508} {\bibfield
		{journal} {\bibinfo  {journal} {Phys. Rev. B}\ }\textbf {\bibinfo {volume}
			{99}},\ \bibinfo {pages} {134508} (\bibinfo {year} {2019})}\BibitemShut
	{NoStop}%
	\bibitem [{\citenamefont {Rasmussen}\ \emph
		{et~al.}(2020{\natexlab{a}})\citenamefont {Rasmussen}, \citenamefont
		{Groenland}, \citenamefont {Gerritsma}, \citenamefont {Schoutens},\ and\
		\citenamefont {Zinner}}]{Rasmussen2020a}%
	\BibitemOpen
	\bibfield  {author} {\bibinfo {author} {\bibfnamefont {S.~E.}\ \bibnamefont
			{Rasmussen}}, \bibinfo {author} {\bibfnamefont {K.}~\bibnamefont
			{Groenland}}, \bibinfo {author} {\bibfnamefont {R.}~\bibnamefont
			{Gerritsma}}, \bibinfo {author} {\bibfnamefont {K.}~\bibnamefont
			{Schoutens}}, \ and\ \bibinfo {author} {\bibfnamefont {N.~T.}\ \bibnamefont
			{Zinner}},\ }\href {\doibase 10.1103/PhysRevA.101.022308} {\bibfield
		{journal} {\bibinfo  {journal} {Phys. Rev. A}\ }\textbf {\bibinfo {volume}
			{101}},\ \bibinfo {pages} {022308} (\bibinfo {year}
		{2020}{\natexlab{a}})}\BibitemShut {NoStop}%
	\bibitem [{\citenamefont {Farhi}\ \emph {et~al.}(2014)\citenamefont {Farhi},
		\citenamefont {Goldstone},\ and\ \citenamefont {Gutmann}}]{Fahri2014}%
	\BibitemOpen
	\bibfield  {author} {\bibinfo {author} {\bibfnamefont {E.}~\bibnamefont
			{Farhi}}, \bibinfo {author} {\bibfnamefont {J.}~\bibnamefont {Goldstone}}, \
		and\ \bibinfo {author} {\bibfnamefont {S.}~\bibnamefont {Gutmann}},\
	}\href@noop {} {\enquote {\bibinfo {title} {A quantum approximate
				optimization algorithm},}\ } (\bibinfo {year} {2014}),\ \bibinfo {note}
	{arXiv:1411.4028}\BibitemShut {NoStop}%
	\bibitem [{\citenamefont {Otterbach}\ \emph {et~al.}(2017)\citenamefont
		{Otterbach}, \citenamefont {Manenti}, \citenamefont {Alidoust}, \citenamefont
		{Bestwick}, \citenamefont {Block}, \citenamefont {Bloom}, \citenamefont
		{Caldwell}, \citenamefont {Didier}, \citenamefont {Fried}, \citenamefont
		{Hong}, \citenamefont {Karalekas}, \citenamefont {Osborn}, \citenamefont
		{Papageorge}, \citenamefont {Peterson}, \citenamefont {Prawiroatmodjo},
		\citenamefont {Rubin}, \citenamefont {Ryan}, \citenamefont {Scarabelli},
		\citenamefont {Scheer}, \citenamefont {Sete}, \citenamefont {Sivarajah},
		\citenamefont {Smith}, \citenamefont {Staley}, \citenamefont {Tezak},
		\citenamefont {Zeng}, \citenamefont {Hudson}, \citenamefont {Johnson},
		\citenamefont {Reagor}, \citenamefont {da~Silva},\ and\ \citenamefont
		{Rigetti}}]{Otterbach2017}%
	\BibitemOpen
	\bibfield  {author} {\bibinfo {author} {\bibfnamefont {J.~S.}\ \bibnamefont
			{Otterbach}}, \bibinfo {author} {\bibfnamefont {R.}~\bibnamefont {Manenti}},
		\bibinfo {author} {\bibfnamefont {N.}~\bibnamefont {Alidoust}}, \bibinfo
		{author} {\bibfnamefont {A.}~\bibnamefont {Bestwick}}, \bibinfo {author}
		{\bibfnamefont {M.}~\bibnamefont {Block}}, \bibinfo {author} {\bibfnamefont
			{B.}~\bibnamefont {Bloom}}, \bibinfo {author} {\bibfnamefont
			{S.}~\bibnamefont {Caldwell}}, \bibinfo {author} {\bibfnamefont
			{N.}~\bibnamefont {Didier}}, \bibinfo {author} {\bibfnamefont {E.~S.}\
			\bibnamefont {Fried}}, \bibinfo {author} {\bibfnamefont {S.}~\bibnamefont
			{Hong}}, \bibinfo {author} {\bibfnamefont {P.}~\bibnamefont {Karalekas}},
		\bibinfo {author} {\bibfnamefont {C.~B.}\ \bibnamefont {Osborn}}, \bibinfo
		{author} {\bibfnamefont {A.}~\bibnamefont {Papageorge}}, \bibinfo {author}
		{\bibfnamefont {E.~C.}\ \bibnamefont {Peterson}}, \bibinfo {author}
		{\bibfnamefont {G.}~\bibnamefont {Prawiroatmodjo}}, \bibinfo {author}
		{\bibfnamefont {N.}~\bibnamefont {Rubin}}, \bibinfo {author} {\bibfnamefont
			{C.~A.}\ \bibnamefont {Ryan}}, \bibinfo {author} {\bibfnamefont
			{D.}~\bibnamefont {Scarabelli}}, \bibinfo {author} {\bibfnamefont
			{M.}~\bibnamefont {Scheer}}, \bibinfo {author} {\bibfnamefont {E.~A.}\
			\bibnamefont {Sete}}, \bibinfo {author} {\bibfnamefont {P.}~\bibnamefont
			{Sivarajah}}, \bibinfo {author} {\bibfnamefont {R.~S.}\ \bibnamefont
			{Smith}}, \bibinfo {author} {\bibfnamefont {A.}~\bibnamefont {Staley}},
		\bibinfo {author} {\bibfnamefont {N.}~\bibnamefont {Tezak}}, \bibinfo
		{author} {\bibfnamefont {W.~J.}\ \bibnamefont {Zeng}}, \bibinfo {author}
		{\bibfnamefont {A.}~\bibnamefont {Hudson}}, \bibinfo {author} {\bibfnamefont
			{B.~R.}\ \bibnamefont {Johnson}}, \bibinfo {author} {\bibfnamefont
			{M.}~\bibnamefont {Reagor}}, \bibinfo {author} {\bibfnamefont {M.~P.}\
			\bibnamefont {da~Silva}}, \ and\ \bibinfo {author} {\bibfnamefont
			{C.}~\bibnamefont {Rigetti}},\ }\href@noop {} {\enquote {\bibinfo {title}
			{Unsupervised machine learning on a hybrid quantum computer},}\ } (\bibinfo
	{year} {2017}),\ \bibinfo {note} {arXiv:1712.05771}\BibitemShut {NoStop}%
	\bibitem [{\citenamefont {Moll}\ \emph {et~al.}(2018)\citenamefont {Moll},
		\citenamefont {Barkoutsos}, \citenamefont {Bishop}, \citenamefont {Chow},
		\citenamefont {Cross}, \citenamefont {Egger}, \citenamefont {Filipp},
		\citenamefont {Fuhrer}, \citenamefont {Gambetta}, \citenamefont {Ganzhorn},
		\citenamefont {Kandala}, \citenamefont {Mezzacapo}, \citenamefont {Muller},
		\citenamefont {Riess}, \citenamefont {Salis}, \citenamefont {Smolin},
		\citenamefont {Tavernelli},\ and\ \citenamefont {Temme}}]{Moll2018}%
	\BibitemOpen
	\bibfield  {author} {\bibinfo {author} {\bibfnamefont {N.}~\bibnamefont
			{Moll}}, \bibinfo {author} {\bibfnamefont {P.}~\bibnamefont {Barkoutsos}},
		\bibinfo {author} {\bibfnamefont {L.~S.}\ \bibnamefont {Bishop}}, \bibinfo
		{author} {\bibfnamefont {J.~M.}\ \bibnamefont {Chow}}, \bibinfo {author}
		{\bibfnamefont {A.}~\bibnamefont {Cross}}, \bibinfo {author} {\bibfnamefont
			{D.~J.}\ \bibnamefont {Egger}}, \bibinfo {author} {\bibfnamefont
			{S.}~\bibnamefont {Filipp}}, \bibinfo {author} {\bibfnamefont
			{A.}~\bibnamefont {Fuhrer}}, \bibinfo {author} {\bibfnamefont {J.~M.}\
			\bibnamefont {Gambetta}}, \bibinfo {author} {\bibfnamefont {M.}~\bibnamefont
			{Ganzhorn}}, \bibinfo {author} {\bibfnamefont {A.}~\bibnamefont {Kandala}},
		\bibinfo {author} {\bibfnamefont {A.}~\bibnamefont {Mezzacapo}}, \bibinfo
		{author} {\bibfnamefont {P.}~\bibnamefont {Muller}}, \bibinfo {author}
		{\bibfnamefont {W.}~\bibnamefont {Riess}}, \bibinfo {author} {\bibfnamefont
			{G.}~\bibnamefont {Salis}}, \bibinfo {author} {\bibfnamefont
			{J.}~\bibnamefont {Smolin}}, \bibinfo {author} {\bibfnamefont
			{I.}~\bibnamefont {Tavernelli}}, \ and\ \bibinfo {author} {\bibfnamefont
			{K.}~\bibnamefont {Temme}},\ }\href {\doibase 10.1088/2058-9565/aab822}
	{\bibfield  {journal} {\bibinfo  {journal} {Quantum Science and Technology}\
		}\textbf {\bibinfo {volume} {3}},\ \bibinfo {pages} {030503} (\bibinfo {year}
		{2018})}\BibitemShut {NoStop}%
	\bibitem [{\citenamefont {Romero}\ \emph {et~al.}(2017)\citenamefont {Romero},
		\citenamefont {Olson},\ and\ \citenamefont {Aspuru-Guzik}}]{Romero2017}%
	\BibitemOpen
	\bibfield  {author} {\bibinfo {author} {\bibfnamefont {J.}~\bibnamefont
			{Romero}}, \bibinfo {author} {\bibfnamefont {J.~P.}\ \bibnamefont {Olson}}, \
		and\ \bibinfo {author} {\bibfnamefont {A.}~\bibnamefont {Aspuru-Guzik}},\
	}\href@noop {} {\bibfield  {journal} {\bibinfo  {journal} {Quantum Sci.
				Technol.}\ }\textbf {\bibinfo {volume} {2}},\ \bibinfo {pages} {045001}
		(\bibinfo {year} {2017})}\BibitemShut {NoStop}%
	\bibitem [{\citenamefont {Johnson}\ \emph {et~al.}(2017)\citenamefont
		{Johnson}, \citenamefont {Romero}, \citenamefont {Olson}, \citenamefont
		{Cao},\ and\ \citenamefont {Aspuru-Guzik}}]{Johnson2017}%
	\BibitemOpen
	\bibfield  {author} {\bibinfo {author} {\bibfnamefont {P.~D.}\ \bibnamefont
			{Johnson}}, \bibinfo {author} {\bibfnamefont {J.}~\bibnamefont {Romero}},
		\bibinfo {author} {\bibfnamefont {J.}~\bibnamefont {Olson}}, \bibinfo
		{author} {\bibfnamefont {Y.}~\bibnamefont {Cao}}, \ and\ \bibinfo {author}
		{\bibfnamefont {A.}~\bibnamefont {Aspuru-Guzik}},\ }\href@noop {} {\enquote
		{\bibinfo {title} {Qvector: an algorithm for device-tailored quantum error
				correction},}\ } (\bibinfo {year} {2017}),\ \bibinfo {note}
	{arXiv:1711.02249}\BibitemShut {NoStop}%
	\bibitem [{\citenamefont {Farhi}\ and\ \citenamefont
		{Neven}(2018)}]{Fahri2018}%
	\BibitemOpen
	\bibfield  {author} {\bibinfo {author} {\bibfnamefont {E.}~\bibnamefont
			{Farhi}}\ and\ \bibinfo {author} {\bibfnamefont {H.}~\bibnamefont {Neven}},\
	}\href@noop {} {\enquote {\bibinfo {title} {Classification with quantum
				neural networks on near term processors},}\ } (\bibinfo {year} {2018}),\
	\bibinfo {note} {arXiv:1802.06002}\BibitemShut {NoStop}%
	\bibitem [{\citenamefont {Havl{\'i}cek}\ \emph {et~al.}(2019)\citenamefont
		{Havl{\'i}cek}, \citenamefont {C{\'o}rcoles}, \citenamefont {Temme},
		\citenamefont {Harrow}, \citenamefont {Kandala}, \citenamefont {Chow},\ and\
		\citenamefont {Gambetta}}]{Havlicek2019}%
	\BibitemOpen
	\bibfield  {author} {\bibinfo {author} {\bibfnamefont {V.}~\bibnamefont
			{Havl{\'i}cek}}, \bibinfo {author} {\bibfnamefont {A.~D.}\ \bibnamefont
			{C{\'o}rcoles}}, \bibinfo {author} {\bibfnamefont {K.}~\bibnamefont {Temme}},
		\bibinfo {author} {\bibfnamefont {A.~W.}\ \bibnamefont {Harrow}}, \bibinfo
		{author} {\bibfnamefont {A.}~\bibnamefont {Kandala}}, \bibinfo {author}
		{\bibfnamefont {J.~M.}\ \bibnamefont {Chow}}, \ and\ \bibinfo {author}
		{\bibfnamefont {J.~M.}\ \bibnamefont {Gambetta}},\ }\href {\doibase
		10.1038/s41586-019-0980-2} {\bibfield  {journal} {\bibinfo  {journal}
			{Nature}\ }\textbf {\bibinfo {volume} {567}},\ \bibinfo {pages} {209}
		(\bibinfo {year} {2019})}\BibitemShut {NoStop}%
	\bibitem [{\citenamefont {Schuld}\ \emph {et~al.}(2020)\citenamefont {Schuld},
		\citenamefont {Bocharov}, \citenamefont {Svore},\ and\ \citenamefont
		{Wiebe}}]{Schuld2020}%
	\BibitemOpen
	\bibfield  {author} {\bibinfo {author} {\bibfnamefont {M.}~\bibnamefont
			{Schuld}}, \bibinfo {author} {\bibfnamefont {A.}~\bibnamefont {Bocharov}},
		\bibinfo {author} {\bibfnamefont {K.~M.}\ \bibnamefont {Svore}}, \ and\
		\bibinfo {author} {\bibfnamefont {N.}~\bibnamefont {Wiebe}},\ }\href
	{\doibase 10.1103/PhysRevA.101.032308} {\bibfield  {journal} {\bibinfo
			{journal} {Phys. Rev. A}\ }\textbf {\bibinfo {volume} {101}},\ \bibinfo
		{pages} {032308} (\bibinfo {year} {2020})}\BibitemShut {NoStop}%
	\bibitem [{\citenamefont {Hubregtsen}\ \emph {et~al.}(2020)\citenamefont
		{Hubregtsen}, \citenamefont {Pichlmeier}, \citenamefont {Stecher},\ and\
		\citenamefont {Bertels}}]{Hubregtsen2020}%
	\BibitemOpen
	\bibfield  {author} {\bibinfo {author} {\bibfnamefont {T.}~\bibnamefont
			{Hubregtsen}}, \bibinfo {author} {\bibfnamefont {J.}~\bibnamefont
			{Pichlmeier}}, \bibinfo {author} {\bibfnamefont {P.}~\bibnamefont {Stecher}},
		\ and\ \bibinfo {author} {\bibfnamefont {K.}~\bibnamefont {Bertels}},\
	}\href@noop {} {\enquote {\bibinfo {title} {Evaluation of parameterized
				quantum circuits: on the design, and the relation between classification
				accuracy, expressibility and entangling capability},}\ } (\bibinfo {year}
	{2020}),\ \bibinfo {note} {arXiv:2003.09887}\BibitemShut {NoStop}%
	\bibitem [{\citenamefont {Dallaire-Demers}\ and\ \citenamefont
		{Killoran}(2018)}]{Dallaire-Demers2018}%
	\BibitemOpen
	\bibfield  {author} {\bibinfo {author} {\bibfnamefont {P.-L.}\ \bibnamefont
			{Dallaire-Demers}}\ and\ \bibinfo {author} {\bibfnamefont {N.}~\bibnamefont
			{Killoran}},\ }\href {\doibase 10.1103/PhysRevA.98.012324} {\bibfield
		{journal} {\bibinfo  {journal} {Phys. Rev. A}\ }\textbf {\bibinfo {volume}
			{98}},\ \bibinfo {pages} {012324} (\bibinfo {year} {2018})}\BibitemShut
	{NoStop}%
	\bibitem [{\citenamefont {Lloyd}\ and\ \citenamefont
		{Weedbrook}(2018)}]{Lloyd2018}%
	\BibitemOpen
	\bibfield  {author} {\bibinfo {author} {\bibfnamefont {S.}~\bibnamefont
			{Lloyd}}\ and\ \bibinfo {author} {\bibfnamefont {C.}~\bibnamefont
			{Weedbrook}},\ }\href {\doibase 10.1103/PhysRevLett.121.040502} {\bibfield
		{journal} {\bibinfo  {journal} {Phys. Rev. Lett.}\ }\textbf {\bibinfo
			{volume} {121}},\ \bibinfo {pages} {040502} (\bibinfo {year}
		{2018})}\BibitemShut {NoStop}%
	\bibitem [{\citenamefont {Zoufal}\ \emph {et~al.}(2019)\citenamefont {Zoufal},
		\citenamefont {Lucchi},\ and\ \citenamefont {Woerner}}]{Zoufal2019}%
	\BibitemOpen
	\bibfield  {author} {\bibinfo {author} {\bibfnamefont {C.}~\bibnamefont
			{Zoufal}}, \bibinfo {author} {\bibfnamefont {A.}~\bibnamefont {Lucchi}}, \
		and\ \bibinfo {author} {\bibfnamefont {S.}~\bibnamefont {Woerner}},\ }\href
	{\doibase 10.1038/s41534-019-0223-2} {\bibfield  {journal} {\bibinfo
			{journal} {npj Quantum Information}\ }\textbf {\bibinfo {volume} {5}},\
		\bibinfo {pages} {103} (\bibinfo {year} {2019})}\BibitemShut {NoStop}%
	\bibitem [{\citenamefont {Zhu}\ \emph {et~al.}(2019)\citenamefont {Zhu},
		\citenamefont {Linke}, \citenamefont {Benedetti}, \citenamefont {Landsman},
		\citenamefont {Nguyen}, \citenamefont {Alderete}, \citenamefont
		{Perdomo-Ortiz}, \citenamefont {Korda}, \citenamefont {Garfoot},
		\citenamefont {Brecque}, \citenamefont {Egan}, \citenamefont {Perdomo},\ and\
		\citenamefont {Monroe}}]{Zhu2019}%
	\BibitemOpen
	\bibfield  {author} {\bibinfo {author} {\bibfnamefont {D.}~\bibnamefont
			{Zhu}}, \bibinfo {author} {\bibfnamefont {N.~M.}\ \bibnamefont {Linke}},
		\bibinfo {author} {\bibfnamefont {M.}~\bibnamefont {Benedetti}}, \bibinfo
		{author} {\bibfnamefont {K.~A.}\ \bibnamefont {Landsman}}, \bibinfo {author}
		{\bibfnamefont {N.~H.}\ \bibnamefont {Nguyen}}, \bibinfo {author}
		{\bibfnamefont {C.~H.}\ \bibnamefont {Alderete}}, \bibinfo {author}
		{\bibfnamefont {A.}~\bibnamefont {Perdomo-Ortiz}}, \bibinfo {author}
		{\bibfnamefont {N.}~\bibnamefont {Korda}}, \bibinfo {author} {\bibfnamefont
			{A.}~\bibnamefont {Garfoot}}, \bibinfo {author} {\bibfnamefont
			{C.}~\bibnamefont {Brecque}}, \bibinfo {author} {\bibfnamefont
			{L.}~\bibnamefont {Egan}}, \bibinfo {author} {\bibfnamefont {O.}~\bibnamefont
			{Perdomo}}, \ and\ \bibinfo {author} {\bibfnamefont {C.}~\bibnamefont
			{Monroe}},\ }\href {\doibase 10.1126/sciadv.aaw9918} {\bibfield  {journal}
		{\bibinfo  {journal} {Science Advances}\ }\textbf {\bibinfo {volume} {5}},\
		\bibinfo {pages} {eaaw9918} (\bibinfo {year} {2019})}\BibitemShut {NoStop}%
	\bibitem [{\citenamefont {Peruzzo}\ \emph {et~al.}(2014)\citenamefont
		{Peruzzo}, \citenamefont {McClean}, \citenamefont {Shadbolt}, \citenamefont
		{Yung}, \citenamefont {Zhou}, \citenamefont {Love}, \citenamefont
		{Aspuru-Guzik},\ and\ \citenamefont {O'Brien}}]{Peruzzo2014}%
	\BibitemOpen
	\bibfield  {author} {\bibinfo {author} {\bibfnamefont {A.}~\bibnamefont
			{Peruzzo}}, \bibinfo {author} {\bibfnamefont {J.}~\bibnamefont {McClean}},
		\bibinfo {author} {\bibfnamefont {P.}~\bibnamefont {Shadbolt}}, \bibinfo
		{author} {\bibfnamefont {M.-H.}\ \bibnamefont {Yung}}, \bibinfo {author}
		{\bibfnamefont {X.-Q.}\ \bibnamefont {Zhou}}, \bibinfo {author}
		{\bibfnamefont {P.~J.}\ \bibnamefont {Love}}, \bibinfo {author}
		{\bibfnamefont {A.}~\bibnamefont {Aspuru-Guzik}}, \ and\ \bibinfo {author}
		{\bibfnamefont {J.~L.}\ \bibnamefont {O'Brien}},\ }\href {\doibase
		10.1038/ncomms5213} {\bibfield  {journal} {\bibinfo  {journal} {Nature
				Communications}\ }\textbf {\bibinfo {volume} {5}},\ \bibinfo {pages} {4213}
		(\bibinfo {year} {2014})}\BibitemShut {NoStop}%
	\bibitem [{\citenamefont {McClean}\ \emph {et~al.}(2016)\citenamefont
		{McClean}, \citenamefont {Romero}, \citenamefont {Babbush},\ and\
		\citenamefont {Aspuru-Guzik}}]{McClean2016}%
	\BibitemOpen
	\bibfield  {author} {\bibinfo {author} {\bibfnamefont {J.~R.}\ \bibnamefont
			{McClean}}, \bibinfo {author} {\bibfnamefont {J.}~\bibnamefont {Romero}},
		\bibinfo {author} {\bibfnamefont {R.}~\bibnamefont {Babbush}}, \ and\
		\bibinfo {author} {\bibfnamefont {A.}~\bibnamefont {Aspuru-Guzik}},\ }\href
	{\doibase 10.1088/1367-2630/18/2/023023} {\bibfield  {journal} {\bibinfo
			{journal} {New Journal of Physics}\ }\textbf {\bibinfo {volume} {18}},\
		\bibinfo {pages} {023023} (\bibinfo {year} {2016})}\BibitemShut {NoStop}%
	\bibitem [{\citenamefont {O'Malley}\ \emph {et~al.}(2016)\citenamefont
		{O'Malley}, \citenamefont {Babbush}, \citenamefont {Kivlichan}, \citenamefont
		{Romero}, \citenamefont {McClean}, \citenamefont {Barends}, \citenamefont
		{Kelly}, \citenamefont {Roushan}, \citenamefont {Tranter}, \citenamefont
		{Ding}, \citenamefont {Campbell}, \citenamefont {Chen}, \citenamefont {Chen},
		\citenamefont {Chiaro}, \citenamefont {Dunsworth}, \citenamefont {Fowler},
		\citenamefont {Jeffrey}, \citenamefont {Lucero}, \citenamefont {Megrant},
		\citenamefont {Mutus}, \citenamefont {Neeley}, \citenamefont {Neill},
		\citenamefont {Quintana}, \citenamefont {Sank}, \citenamefont {Vainsencher},
		\citenamefont {Wenner}, \citenamefont {White}, \citenamefont {Coveney},
		\citenamefont {Love}, \citenamefont {Neven}, \citenamefont {Aspuru-Guzik},\
		and\ \citenamefont {Martinis}}]{Omalley2016}%
	\BibitemOpen
	\bibfield  {author} {\bibinfo {author} {\bibfnamefont {P.~J.~J.}\
			\bibnamefont {O'Malley}}, \bibinfo {author} {\bibfnamefont {R.}~\bibnamefont
			{Babbush}}, \bibinfo {author} {\bibfnamefont {I.~D.}\ \bibnamefont
			{Kivlichan}}, \bibinfo {author} {\bibfnamefont {J.}~\bibnamefont {Romero}},
		\bibinfo {author} {\bibfnamefont {J.~R.}\ \bibnamefont {McClean}}, \bibinfo
		{author} {\bibfnamefont {R.}~\bibnamefont {Barends}}, \bibinfo {author}
		{\bibfnamefont {J.}~\bibnamefont {Kelly}}, \bibinfo {author} {\bibfnamefont
			{P.}~\bibnamefont {Roushan}}, \bibinfo {author} {\bibfnamefont
			{A.}~\bibnamefont {Tranter}}, \bibinfo {author} {\bibfnamefont
			{N.}~\bibnamefont {Ding}}, \bibinfo {author} {\bibfnamefont {B.}~\bibnamefont
			{Campbell}}, \bibinfo {author} {\bibfnamefont {Y.}~\bibnamefont {Chen}},
		\bibinfo {author} {\bibfnamefont {Z.}~\bibnamefont {Chen}}, \bibinfo {author}
		{\bibfnamefont {B.}~\bibnamefont {Chiaro}}, \bibinfo {author} {\bibfnamefont
			{A.}~\bibnamefont {Dunsworth}}, \bibinfo {author} {\bibfnamefont {A.~G.}\
			\bibnamefont {Fowler}}, \bibinfo {author} {\bibfnamefont {E.}~\bibnamefont
			{Jeffrey}}, \bibinfo {author} {\bibfnamefont {E.}~\bibnamefont {Lucero}},
		\bibinfo {author} {\bibfnamefont {A.}~\bibnamefont {Megrant}}, \bibinfo
		{author} {\bibfnamefont {J.~Y.}\ \bibnamefont {Mutus}}, \bibinfo {author}
		{\bibfnamefont {M.}~\bibnamefont {Neeley}}, \bibinfo {author} {\bibfnamefont
			{C.}~\bibnamefont {Neill}}, \bibinfo {author} {\bibfnamefont
			{C.}~\bibnamefont {Quintana}}, \bibinfo {author} {\bibfnamefont
			{D.}~\bibnamefont {Sank}}, \bibinfo {author} {\bibfnamefont {A.}~\bibnamefont
			{Vainsencher}}, \bibinfo {author} {\bibfnamefont {J.}~\bibnamefont {Wenner}},
		\bibinfo {author} {\bibfnamefont {T.~C.}\ \bibnamefont {White}}, \bibinfo
		{author} {\bibfnamefont {P.~V.}\ \bibnamefont {Coveney}}, \bibinfo {author}
		{\bibfnamefont {P.~J.}\ \bibnamefont {Love}}, \bibinfo {author}
		{\bibfnamefont {H.}~\bibnamefont {Neven}}, \bibinfo {author} {\bibfnamefont
			{A.}~\bibnamefont {Aspuru-Guzik}}, \ and\ \bibinfo {author} {\bibfnamefont
			{J.~M.}\ \bibnamefont {Martinis}},\ }\href {\doibase
		10.1103/PhysRevX.6.031007} {\bibfield  {journal} {\bibinfo  {journal} {Phys.
				Rev. X}\ }\textbf {\bibinfo {volume} {6}},\ \bibinfo {pages} {031007}
		(\bibinfo {year} {2016})}\BibitemShut {NoStop}%
	\bibitem [{\citenamefont {Kandala}\ \emph {et~al.}(2017)\citenamefont
		{Kandala}, \citenamefont {Mezzacapo}, \citenamefont {Temme}, \citenamefont
		{Takita}, \citenamefont {Brink}, \citenamefont {Chow},\ and\ \citenamefont
		{Gambetta}}]{Kandala2017}%
	\BibitemOpen
	\bibfield  {author} {\bibinfo {author} {\bibfnamefont {A.}~\bibnamefont
			{Kandala}}, \bibinfo {author} {\bibfnamefont {A.}~\bibnamefont {Mezzacapo}},
		\bibinfo {author} {\bibfnamefont {K.}~\bibnamefont {Temme}}, \bibinfo
		{author} {\bibfnamefont {M.}~\bibnamefont {Takita}}, \bibinfo {author}
		{\bibfnamefont {M.}~\bibnamefont {Brink}}, \bibinfo {author} {\bibfnamefont
			{J.~M.}\ \bibnamefont {Chow}}, \ and\ \bibinfo {author} {\bibfnamefont
			{J.~M.}\ \bibnamefont {Gambetta}},\ }\href {\doibase 10.1038/nature23879}
	{\bibfield  {journal} {\bibinfo  {journal} {Nature}\ }\textbf {\bibinfo
			{volume} {549}},\ \bibinfo {pages} {242} (\bibinfo {year}
		{2017})}\BibitemShut {NoStop}%
	\bibitem [{\citenamefont {Cao}\ \emph {et~al.}(2019)\citenamefont {Cao},
		\citenamefont {Romero}, \citenamefont {Olson}, \citenamefont {Degroote},
		\citenamefont {Johnson}, \citenamefont {Kieferov{\'{a}}}, \citenamefont
		{Kivlichan}, \citenamefont {Menke}, \citenamefont {Peropadre}, \citenamefont
		{Sawaya}, \citenamefont {Sim}, \citenamefont {Veis},\ and\ \citenamefont
		{Aspuru-Guzik}}]{Cao2019}%
	\BibitemOpen
	\bibfield  {author} {\bibinfo {author} {\bibfnamefont {Y.}~\bibnamefont
			{Cao}}, \bibinfo {author} {\bibfnamefont {J.}~\bibnamefont {Romero}},
		\bibinfo {author} {\bibfnamefont {J.~P.}\ \bibnamefont {Olson}}, \bibinfo
		{author} {\bibfnamefont {M.}~\bibnamefont {Degroote}}, \bibinfo {author}
		{\bibfnamefont {P.~D.}\ \bibnamefont {Johnson}}, \bibinfo {author}
		{\bibfnamefont {M.}~\bibnamefont {Kieferov{\'{a}}}}, \bibinfo {author}
		{\bibfnamefont {I.~D.}\ \bibnamefont {Kivlichan}}, \bibinfo {author}
		{\bibfnamefont {T.}~\bibnamefont {Menke}}, \bibinfo {author} {\bibfnamefont
			{B.}~\bibnamefont {Peropadre}}, \bibinfo {author} {\bibfnamefont {N.~P.~D.}\
			\bibnamefont {Sawaya}}, \bibinfo {author} {\bibfnamefont {S.}~\bibnamefont
			{Sim}}, \bibinfo {author} {\bibfnamefont {L.}~\bibnamefont {Veis}}, \ and\
		\bibinfo {author} {\bibfnamefont {A.}~\bibnamefont {Aspuru-Guzik}},\ }\href
	{\doibase 10.1021/acs.chemrev.8b00803} {\bibfield  {journal} {\bibinfo
			{journal} {Chemical Reviews}\ }\textbf {\bibinfo {volume} {119}},\ \bibinfo
		{pages} {10856} (\bibinfo {year} {2019})}\BibitemShut {NoStop}%
	\bibitem [{\citenamefont {Barkoutsos}\ \emph {et~al.}(2018)\citenamefont
		{Barkoutsos}, \citenamefont {Gonthier}, \citenamefont {Sokolov},
		\citenamefont {Moll}, \citenamefont {Salis}, \citenamefont {Fuhrer},
		\citenamefont {Ganzhorn}, \citenamefont {Egger}, \citenamefont {Troyer},
		\citenamefont {Mezzacapo}, \citenamefont {Filipp},\ and\ \citenamefont
		{Tavernelli}}]{Barkoutos2018}%
	\BibitemOpen
	\bibfield  {author} {\bibinfo {author} {\bibfnamefont {P.~K.}\ \bibnamefont
			{Barkoutsos}}, \bibinfo {author} {\bibfnamefont {J.~F.}\ \bibnamefont
			{Gonthier}}, \bibinfo {author} {\bibfnamefont {I.}~\bibnamefont {Sokolov}},
		\bibinfo {author} {\bibfnamefont {N.}~\bibnamefont {Moll}}, \bibinfo {author}
		{\bibfnamefont {G.}~\bibnamefont {Salis}}, \bibinfo {author} {\bibfnamefont
			{A.}~\bibnamefont {Fuhrer}}, \bibinfo {author} {\bibfnamefont
			{M.}~\bibnamefont {Ganzhorn}}, \bibinfo {author} {\bibfnamefont {D.~J.}\
			\bibnamefont {Egger}}, \bibinfo {author} {\bibfnamefont {M.}~\bibnamefont
			{Troyer}}, \bibinfo {author} {\bibfnamefont {A.}~\bibnamefont {Mezzacapo}},
		\bibinfo {author} {\bibfnamefont {S.}~\bibnamefont {Filipp}}, \ and\ \bibinfo
		{author} {\bibfnamefont {I.}~\bibnamefont {Tavernelli}},\ }\href {\doibase
		10.1103/PhysRevA.98.022322} {\bibfield  {journal} {\bibinfo  {journal} {Phys.
				Rev. A}\ }\textbf {\bibinfo {volume} {98}},\ \bibinfo {pages} {022322}
		(\bibinfo {year} {2018})}\BibitemShut {NoStop}%
	\bibitem [{\citenamefont {McCaskey}\ \emph {et~al.}(2019)\citenamefont
		{McCaskey}, \citenamefont {Parks}, \citenamefont {Jakowski}, \citenamefont
		{Moore}, \citenamefont {Morris}, \citenamefont {Humble},\ and\ \citenamefont
		{Pooser}}]{McCaskey2019}%
	\BibitemOpen
	\bibfield  {author} {\bibinfo {author} {\bibfnamefont {A.~J.}\ \bibnamefont
			{McCaskey}}, \bibinfo {author} {\bibfnamefont {Z.~P.}\ \bibnamefont {Parks}},
		\bibinfo {author} {\bibfnamefont {J.}~\bibnamefont {Jakowski}}, \bibinfo
		{author} {\bibfnamefont {S.~V.}\ \bibnamefont {Moore}}, \bibinfo {author}
		{\bibfnamefont {T.~D.}\ \bibnamefont {Morris}}, \bibinfo {author}
		{\bibfnamefont {T.~S.}\ \bibnamefont {Humble}}, \ and\ \bibinfo {author}
		{\bibfnamefont {R.~C.}\ \bibnamefont {Pooser}},\ }\href {\doibase
		10.1038/s41534-019-0209-0} {\bibfield  {journal} {\bibinfo  {journal} {npj
				Quantum Information}\ }\textbf {\bibinfo {volume} {5}},\ \bibinfo {pages}
		{99} (\bibinfo {year} {2019})}\BibitemShut {NoStop}%
	\bibitem [{\citenamefont {Gard}\ \emph {et~al.}(2020)\citenamefont {Gard},
		\citenamefont {Zhu}, \citenamefont {Barron}, \citenamefont {Mayhall},
		\citenamefont {Economou},\ and\ \citenamefont {Barnes}}]{Gard2020}%
	\BibitemOpen
	\bibfield  {author} {\bibinfo {author} {\bibfnamefont {B.~T.}\ \bibnamefont
			{Gard}}, \bibinfo {author} {\bibfnamefont {L.}~\bibnamefont {Zhu}}, \bibinfo
		{author} {\bibfnamefont {G.~S.}\ \bibnamefont {Barron}}, \bibinfo {author}
		{\bibfnamefont {N.~J.}\ \bibnamefont {Mayhall}}, \bibinfo {author}
		{\bibfnamefont {S.~E.}\ \bibnamefont {Economou}}, \ and\ \bibinfo {author}
		{\bibfnamefont {E.}~\bibnamefont {Barnes}},\ }\href {\doibase
		10.1038/s41534-019-0240-1} {\bibfield  {journal} {\bibinfo  {journal} {npj
				Quantum Information}\ }\textbf {\bibinfo {volume} {6}},\ \bibinfo {pages}
		{10} (\bibinfo {year} {2020})}\BibitemShut {NoStop}%
	\bibitem [{\citenamefont {Sim}\ \emph {et~al.}(2019)\citenamefont {Sim},
		\citenamefont {Johnson},\ and\ \citenamefont {Aspuru-Guzik}}]{Sim2019}%
	\BibitemOpen
	\bibfield  {author} {\bibinfo {author} {\bibfnamefont {S.}~\bibnamefont
			{Sim}}, \bibinfo {author} {\bibfnamefont {P.~D.}\ \bibnamefont {Johnson}}, \
		and\ \bibinfo {author} {\bibfnamefont {A.}~\bibnamefont {Aspuru-Guzik}},\
	}\href {\doibase 10.1002/qute.201900070} {\bibfield  {journal} {\bibinfo
			{journal} {Advanced Quantum Technologies}\ }\textbf {\bibinfo {volume} {2}},\
		\bibinfo {pages} {1900070} (\bibinfo {year} {2019})}\BibitemShut {NoStop}%
	\bibitem [{\citenamefont {Geller}(2018)}]{Geller2018}%
	\BibitemOpen
	\bibfield  {author} {\bibinfo {author} {\bibfnamefont {M.~R.}\ \bibnamefont
			{Geller}},\ }\href {\doibase 10.1103/PhysRevApplied.10.024052} {\bibfield
		{journal} {\bibinfo  {journal} {Phys. Rev. Applied}\ }\textbf {\bibinfo
			{volume} {10}},\ \bibinfo {pages} {024052} (\bibinfo {year}
		{2018})}\BibitemShut {NoStop}%
	\bibitem [{\citenamefont {Du}\ \emph {et~al.}(2020)\citenamefont {Du},
		\citenamefont {Hsieh}, \citenamefont {Liu},\ and\ \citenamefont
		{Tao}}]{Du2018}%
	\BibitemOpen
	\bibfield  {author} {\bibinfo {author} {\bibfnamefont {Y.}~\bibnamefont
			{Du}}, \bibinfo {author} {\bibfnamefont {M.-H.}\ \bibnamefont {Hsieh}},
		\bibinfo {author} {\bibfnamefont {T.}~\bibnamefont {Liu}}, \ and\ \bibinfo
		{author} {\bibfnamefont {D.}~\bibnamefont {Tao}},\ }\href {\doibase
		10.1103/PhysRevResearch.2.033125} {\bibfield  {journal} {\bibinfo  {journal}
			{Phys. Rev. Research}\ }\textbf {\bibinfo {volume} {2}},\ \bibinfo {pages}
		{033125} (\bibinfo {year} {2020})}\BibitemShut {NoStop}%
	\bibitem [{\citenamefont {Benedetti}\ \emph {et~al.}(2019)\citenamefont
		{Benedetti}, \citenamefont {Lloyd}, \citenamefont {Sack},\ and\ \citenamefont
		{Fiorentini}}]{Benedetti2019}%
	\BibitemOpen
	\bibfield  {author} {\bibinfo {author} {\bibfnamefont {M.}~\bibnamefont
			{Benedetti}}, \bibinfo {author} {\bibfnamefont {E.}~\bibnamefont {Lloyd}},
		\bibinfo {author} {\bibfnamefont {S.}~\bibnamefont {Sack}}, \ and\ \bibinfo
		{author} {\bibfnamefont {M.}~\bibnamefont {Fiorentini}},\ }\href {\doibase
		10.1088/2058-9565/ab4eb5} {\bibfield  {journal} {\bibinfo  {journal} {Quantum
				Science and Technology}\ }\textbf {\bibinfo {volume} {4}},\ \bibinfo {pages}
		{043001} (\bibinfo {year} {2019})}\BibitemShut {NoStop}%
	\bibitem [{\citenamefont {Mitarai}\ \emph {et~al.}(2018)\citenamefont
		{Mitarai}, \citenamefont {Negoro}, \citenamefont {Kitagawa},\ and\
		\citenamefont {Fujii}}]{Mitarai2018}%
	\BibitemOpen
	\bibfield  {author} {\bibinfo {author} {\bibfnamefont {K.}~\bibnamefont
			{Mitarai}}, \bibinfo {author} {\bibfnamefont {M.}~\bibnamefont {Negoro}},
		\bibinfo {author} {\bibfnamefont {M.}~\bibnamefont {Kitagawa}}, \ and\
		\bibinfo {author} {\bibfnamefont {K.}~\bibnamefont {Fujii}},\ }\href
	{\doibase 10.1103/PhysRevA.98.032309} {\bibfield  {journal} {\bibinfo
			{journal} {Phys. Rev. A}\ }\textbf {\bibinfo {volume} {98}},\ \bibinfo
		{pages} {032309} (\bibinfo {year} {2018})}\BibitemShut {NoStop}%
	\bibitem [{\citenamefont {Rasmussen}\ \emph
		{et~al.}(2020{\natexlab{b}})\citenamefont {Rasmussen}, \citenamefont {Loft},
		\citenamefont {B{\ae}kkegaard}, \citenamefont {Kues},\ and\ \citenamefont
		{Zinner}}]{Rasmussen2020c}%
	\BibitemOpen
	\bibfield  {author} {\bibinfo {author} {\bibfnamefont {S.~E.}\ \bibnamefont
			{Rasmussen}}, \bibinfo {author} {\bibfnamefont {N.~J.~S.}\ \bibnamefont
			{Loft}}, \bibinfo {author} {\bibfnamefont {T.}~\bibnamefont
			{B{\ae}kkegaard}}, \bibinfo {author} {\bibfnamefont {M.}~\bibnamefont
			{Kues}}, \ and\ \bibinfo {author} {\bibfnamefont {N.~T.}\ \bibnamefont
			{Zinner}},\ }\href {\doibase https://doi.org/10.1002/qute.202000063}
	{\bibfield  {journal} {\bibinfo  {journal} {Advanced Quantum Technologies}\
		}\textbf {\bibinfo {volume} {3}},\ \bibinfo {pages} {2000063} (\bibinfo
		{year} {2020}{\natexlab{b}})}\BibitemShut {NoStop}%
	\bibitem [{\citenamefont {Koch}\ \emph {et~al.}(2007)\citenamefont {Koch},
		\citenamefont {Yu}, \citenamefont {Gambetta}, \citenamefont {Houck},
		\citenamefont {Schuster}, \citenamefont {Majer}, \citenamefont {Blais},
		\citenamefont {Devoret}, \citenamefont {Girvin},\ and\ \citenamefont
		{Schoelkopf}}]{Koch2007}%
	\BibitemOpen
	\bibfield  {author} {\bibinfo {author} {\bibfnamefont {J.}~\bibnamefont
			{Koch}}, \bibinfo {author} {\bibfnamefont {T.~M.}\ \bibnamefont {Yu}},
		\bibinfo {author} {\bibfnamefont {J.}~\bibnamefont {Gambetta}}, \bibinfo
		{author} {\bibfnamefont {A.~A.}\ \bibnamefont {Houck}}, \bibinfo {author}
		{\bibfnamefont {D.~I.}\ \bibnamefont {Schuster}}, \bibinfo {author}
		{\bibfnamefont {J.}~\bibnamefont {Majer}}, \bibinfo {author} {\bibfnamefont
			{A.}~\bibnamefont {Blais}}, \bibinfo {author} {\bibfnamefont {M.~H.}\
			\bibnamefont {Devoret}}, \bibinfo {author} {\bibfnamefont {S.~M.}\
			\bibnamefont {Girvin}}, \ and\ \bibinfo {author} {\bibfnamefont {R.~J.}\
			\bibnamefont {Schoelkopf}},\ }\href {\doibase 10.1103/PhysRevA.76.042319}
	{\bibfield  {journal} {\bibinfo  {journal} {Phys. Rev. A}\ }\textbf {\bibinfo
			{volume} {76}},\ \bibinfo {pages} {042319} (\bibinfo {year}
		{2007})}\BibitemShut {NoStop}%
	\bibitem [{\citenamefont {Schreier}\ \emph {et~al.}(2008)\citenamefont
		{Schreier}, \citenamefont {Houck}, \citenamefont {Koch}, \citenamefont
		{Schuster}, \citenamefont {Johnson}, \citenamefont {Chow}, \citenamefont
		{Gambetta}, \citenamefont {Majer}, \citenamefont {Frunzio}, \citenamefont
		{Devoret}, \citenamefont {Girvin},\ and\ \citenamefont
		{Schoelkopf}}]{Schreier2008}%
	\BibitemOpen
	\bibfield  {author} {\bibinfo {author} {\bibfnamefont {J.~A.}\ \bibnamefont
			{Schreier}}, \bibinfo {author} {\bibfnamefont {A.~A.}\ \bibnamefont {Houck}},
		\bibinfo {author} {\bibfnamefont {J.}~\bibnamefont {Koch}}, \bibinfo {author}
		{\bibfnamefont {D.~I.}\ \bibnamefont {Schuster}}, \bibinfo {author}
		{\bibfnamefont {B.~R.}\ \bibnamefont {Johnson}}, \bibinfo {author}
		{\bibfnamefont {J.~M.}\ \bibnamefont {Chow}}, \bibinfo {author}
		{\bibfnamefont {J.~M.}\ \bibnamefont {Gambetta}}, \bibinfo {author}
		{\bibfnamefont {J.}~\bibnamefont {Majer}}, \bibinfo {author} {\bibfnamefont
			{L.}~\bibnamefont {Frunzio}}, \bibinfo {author} {\bibfnamefont {M.~H.}\
			\bibnamefont {Devoret}}, \bibinfo {author} {\bibfnamefont {S.~M.}\
			\bibnamefont {Girvin}}, \ and\ \bibinfo {author} {\bibfnamefont {R.~J.}\
			\bibnamefont {Schoelkopf}},\ }\href {\doibase 10.1103/PhysRevB.77.180502}
	{\bibfield  {journal} {\bibinfo  {journal} {Phys. Rev. B}\ }\textbf {\bibinfo
			{volume} {77}},\ \bibinfo {pages} {180502} (\bibinfo {year}
		{2008})}\BibitemShut {NoStop}%
	\bibitem [{\citenamefont {Schuch}\ and\ \citenamefont
		{Siewert}(2003)}]{Schuch2003}%
	\BibitemOpen
	\bibfield  {author} {\bibinfo {author} {\bibfnamefont {N.}~\bibnamefont
			{Schuch}}\ and\ \bibinfo {author} {\bibfnamefont {J.}~\bibnamefont
			{Siewert}},\ }\href {\doibase 10.1103/PhysRevA.67.032301} {\bibfield
		{journal} {\bibinfo  {journal} {Phys. Rev. A}\ }\textbf {\bibinfo {volume}
			{67}},\ \bibinfo {pages} {032301} (\bibinfo {year} {2003})}\BibitemShut
	{NoStop}%
	\bibitem [{\citenamefont {Tanamoto}\ \emph {et~al.}(2008)\citenamefont
		{Tanamoto}, \citenamefont {Maruyama}, \citenamefont {Liu}, \citenamefont
		{Hu},\ and\ \citenamefont {Nori}}]{Tanamoto2008}%
	\BibitemOpen
	\bibfield  {author} {\bibinfo {author} {\bibfnamefont {T.}~\bibnamefont
			{Tanamoto}}, \bibinfo {author} {\bibfnamefont {K.}~\bibnamefont {Maruyama}},
		\bibinfo {author} {\bibfnamefont {Y.-x.}\ \bibnamefont {Liu}}, \bibinfo
		{author} {\bibfnamefont {X.}~\bibnamefont {Hu}}, \ and\ \bibinfo {author}
		{\bibfnamefont {F.}~\bibnamefont {Nori}},\ }\href {\doibase
		10.1103/PhysRevA.78.062313} {\bibfield  {journal} {\bibinfo  {journal} {Phys.
				Rev. A}\ }\textbf {\bibinfo {volume} {78}},\ \bibinfo {pages} {062313}
		(\bibinfo {year} {2008})}\BibitemShut {NoStop}%
	\bibitem [{\citenamefont {Tanamoto}\ \emph {et~al.}(2009)\citenamefont
		{Tanamoto}, \citenamefont {Liu}, \citenamefont {Hu},\ and\ \citenamefont
		{Nori}}]{Tanamoto2009}%
	\BibitemOpen
	\bibfield  {author} {\bibinfo {author} {\bibfnamefont {T.}~\bibnamefont
			{Tanamoto}}, \bibinfo {author} {\bibfnamefont {Y.-x.}\ \bibnamefont {Liu}},
		\bibinfo {author} {\bibfnamefont {X.}~\bibnamefont {Hu}}, \ and\ \bibinfo
		{author} {\bibfnamefont {F.}~\bibnamefont {Nori}},\ }\href {\doibase
		10.1103/PhysRevLett.102.100501} {\bibfield  {journal} {\bibinfo  {journal}
			{Phys. Rev. Lett.}\ }\textbf {\bibinfo {volume} {102}},\ \bibinfo {pages}
		{100501} (\bibinfo {year} {2009})}\BibitemShut {NoStop}%
	\bibitem [{\citenamefont {You}\ and\ \citenamefont {Nori}(2005)}]{You2005}%
	\BibitemOpen
	\bibfield  {author} {\bibinfo {author} {\bibfnamefont {J.~Q.}\ \bibnamefont
			{You}}\ and\ \bibinfo {author} {\bibfnamefont {F.}~\bibnamefont {Nori}},\
	}\href {\doibase 10.1063/1.2155757} {\bibfield  {journal} {\bibinfo
			{journal} {Physics Today}\ }\textbf {\bibinfo {volume} {58}},\ \bibinfo
		{pages} {42} (\bibinfo {year} {2005})},\ \Eprint
	{http://arxiv.org/abs/https://doi.org/10.1063/1.2155757}
	{https://doi.org/10.1063/1.2155757} \BibitemShut {NoStop}%
	\bibitem [{\citenamefont {Zagoskin}\ \emph {et~al.}(2006)\citenamefont
		{Zagoskin}, \citenamefont {Ashhab}, \citenamefont {Johansson},\ and\
		\citenamefont {Nori}}]{Zagoskin2006}%
	\BibitemOpen
	\bibfield  {author} {\bibinfo {author} {\bibfnamefont {A.~M.}\ \bibnamefont
			{Zagoskin}}, \bibinfo {author} {\bibfnamefont {S.}~\bibnamefont {Ashhab}},
		\bibinfo {author} {\bibfnamefont {J.~R.}\ \bibnamefont {Johansson}}, \ and\
		\bibinfo {author} {\bibfnamefont {F.}~\bibnamefont {Nori}},\ }\href {\doibase
		10.1103/PhysRevLett.97.077001} {\bibfield  {journal} {\bibinfo  {journal}
			{Phys. Rev. Lett.}\ }\textbf {\bibinfo {volume} {97}},\ \bibinfo {pages}
		{077001} (\bibinfo {year} {2006})}\BibitemShut {NoStop}%
	\bibitem [{\citenamefont {McKay}\ \emph {et~al.}(2016)\citenamefont {McKay},
		\citenamefont {Filipp}, \citenamefont {Mezzacapo}, \citenamefont {Magesan},
		\citenamefont {Chow},\ and\ \citenamefont {Gambetta}}]{McKay2016}%
	\BibitemOpen
	\bibfield  {author} {\bibinfo {author} {\bibfnamefont {D.~C.}\ \bibnamefont
			{McKay}}, \bibinfo {author} {\bibfnamefont {S.}~\bibnamefont {Filipp}},
		\bibinfo {author} {\bibfnamefont {A.}~\bibnamefont {Mezzacapo}}, \bibinfo
		{author} {\bibfnamefont {E.}~\bibnamefont {Magesan}}, \bibinfo {author}
		{\bibfnamefont {J.~M.}\ \bibnamefont {Chow}}, \ and\ \bibinfo {author}
		{\bibfnamefont {J.~M.}\ \bibnamefont {Gambetta}},\ }\href {\doibase
		10.1103/PhysRevApplied.6.064007} {\bibfield  {journal} {\bibinfo  {journal}
			{Phys. Rev. Applied}\ }\textbf {\bibinfo {volume} {6}},\ \bibinfo {pages}
		{064007} (\bibinfo {year} {2016})}\BibitemShut {NoStop}%
	\bibitem [{\citenamefont {Dewes}\ \emph {et~al.}(2012)\citenamefont {Dewes},
		\citenamefont {Ong}, \citenamefont {Schmitt}, \citenamefont {Lauro},
		\citenamefont {Boulant}, \citenamefont {Bertet}, \citenamefont {Vion},\ and\
		\citenamefont {Esteve}}]{Dewes2012}%
	\BibitemOpen
	\bibfield  {author} {\bibinfo {author} {\bibfnamefont {A.}~\bibnamefont
			{Dewes}}, \bibinfo {author} {\bibfnamefont {F.~R.}\ \bibnamefont {Ong}},
		\bibinfo {author} {\bibfnamefont {V.}~\bibnamefont {Schmitt}}, \bibinfo
		{author} {\bibfnamefont {R.}~\bibnamefont {Lauro}}, \bibinfo {author}
		{\bibfnamefont {N.}~\bibnamefont {Boulant}}, \bibinfo {author} {\bibfnamefont
			{P.}~\bibnamefont {Bertet}}, \bibinfo {author} {\bibfnamefont
			{D.}~\bibnamefont {Vion}}, \ and\ \bibinfo {author} {\bibfnamefont
			{D.}~\bibnamefont {Esteve}},\ }\href {\doibase
		10.1103/PhysRevLett.108.057002} {\bibfield  {journal} {\bibinfo  {journal}
			{Phys. Rev. Lett.}\ }\textbf {\bibinfo {volume} {108}},\ \bibinfo {pages}
		{057002} (\bibinfo {year} {2012})}\BibitemShut {NoStop}%
	\bibitem [{\citenamefont {Salath\'e}\ \emph {et~al.}(2015)\citenamefont
		{Salath\'e}, \citenamefont {Mondal}, \citenamefont {Oppliger}, \citenamefont
		{Heinsoo}, \citenamefont {Kurpiers}, \citenamefont
		{Poto\ifmmode~\check{c}\else \v{c}\fi{}nik}, \citenamefont {Mezzacapo},
		\citenamefont {Las~Heras}, \citenamefont {Lamata}, \citenamefont {Solano},
		\citenamefont {Filipp},\ and\ \citenamefont {Wallraff}}]{Salathe2015}%
	\BibitemOpen
	\bibfield  {author} {\bibinfo {author} {\bibfnamefont {Y.}~\bibnamefont
			{Salath\'e}}, \bibinfo {author} {\bibfnamefont {M.}~\bibnamefont {Mondal}},
		\bibinfo {author} {\bibfnamefont {M.}~\bibnamefont {Oppliger}}, \bibinfo
		{author} {\bibfnamefont {J.}~\bibnamefont {Heinsoo}}, \bibinfo {author}
		{\bibfnamefont {P.}~\bibnamefont {Kurpiers}}, \bibinfo {author}
		{\bibfnamefont {A.}~\bibnamefont {Poto\ifmmode~\check{c}\else
				\v{c}\fi{}nik}}, \bibinfo {author} {\bibfnamefont {A.}~\bibnamefont
			{Mezzacapo}}, \bibinfo {author} {\bibfnamefont {U.}~\bibnamefont
			{Las~Heras}}, \bibinfo {author} {\bibfnamefont {L.}~\bibnamefont {Lamata}},
		\bibinfo {author} {\bibfnamefont {E.}~\bibnamefont {Solano}}, \bibinfo
		{author} {\bibfnamefont {S.}~\bibnamefont {Filipp}}, \ and\ \bibinfo {author}
		{\bibfnamefont {A.}~\bibnamefont {Wallraff}},\ }\href {\doibase
		10.1103/PhysRevX.5.021027} {\bibfield  {journal} {\bibinfo  {journal} {Phys.
				Rev. X}\ }\textbf {\bibinfo {volume} {5}},\ \bibinfo {pages} {021027}
		(\bibinfo {year} {2015})}\BibitemShut {NoStop}%
	\bibitem [{\citenamefont {Vool}\ and\ \citenamefont
		{Devoret}(2017)}]{Vool2017}%
	\BibitemOpen
	\bibfield  {author} {\bibinfo {author} {\bibfnamefont {U.}~\bibnamefont
			{Vool}}\ and\ \bibinfo {author} {\bibfnamefont {M.}~\bibnamefont {Devoret}},\
	}\href {\doibase https://doi.org/10.1002/cta.2359} {\bibfield  {journal}
		{\bibinfo  {journal} {International Journal of Circuit Theory and
				Applications}\ }\textbf {\bibinfo {volume} {45}},\ \bibinfo {pages} {897}
		(\bibinfo {year} {2017})}\BibitemShut {NoStop}%
	\bibitem [{\citenamefont {Krantz}\ \emph {et~al.}(2019)\citenamefont {Krantz},
		\citenamefont {Kjaergaard}, \citenamefont {Yan}, \citenamefont {Orlando},
		\citenamefont {Gustavsson},\ and\ \citenamefont {Oliver}}]{Krantz2019}%
	\BibitemOpen
	\bibfield  {author} {\bibinfo {author} {\bibfnamefont {P.}~\bibnamefont
			{Krantz}}, \bibinfo {author} {\bibfnamefont {M.}~\bibnamefont {Kjaergaard}},
		\bibinfo {author} {\bibfnamefont {F.}~\bibnamefont {Yan}}, \bibinfo {author}
		{\bibfnamefont {T.~P.}\ \bibnamefont {Orlando}}, \bibinfo {author}
		{\bibfnamefont {S.}~\bibnamefont {Gustavsson}}, \ and\ \bibinfo {author}
		{\bibfnamefont {W.~D.}\ \bibnamefont {Oliver}},\ }\href {\doibase
		10.1063/1.5089550} {\bibfield  {journal} {\bibinfo  {journal} {Applied
				Physics Reviews}\ }\textbf {\bibinfo {volume} {6}},\ \bibinfo {pages}
		{021318} (\bibinfo {year} {2019})}\BibitemShut {NoStop}%
	\bibitem [{\citenamefont {Rasmussen}\ \emph {et~al.}(2021)\citenamefont
		{Rasmussen}, \citenamefont {Christensen}, \citenamefont {Pedersen},
		\citenamefont {Kristensen}, \citenamefont {B\ae{}kkegaard}, \citenamefont
		{Loft},\ and\ \citenamefont {Zinner}}]{Rasmussen2021}%
	\BibitemOpen
	\bibfield  {author} {\bibinfo {author} {\bibfnamefont {S.}~\bibnamefont
			{Rasmussen}}, \bibinfo {author} {\bibfnamefont {K.}~\bibnamefont
			{Christensen}}, \bibinfo {author} {\bibfnamefont {S.}~\bibnamefont
			{Pedersen}}, \bibinfo {author} {\bibfnamefont {L.}~\bibnamefont
			{Kristensen}}, \bibinfo {author} {\bibfnamefont {T.}~\bibnamefont
			{B\ae{}kkegaard}}, \bibinfo {author} {\bibfnamefont {N.}~\bibnamefont
			{Loft}}, \ and\ \bibinfo {author} {\bibfnamefont {N.}~\bibnamefont
			{Zinner}},\ }\href {\doibase 10.1103/PRXQuantum.2.040204} {\bibfield
		{journal} {\bibinfo  {journal} {PRX Quantum}\ }\textbf {\bibinfo {volume}
			{2}},\ \bibinfo {pages} {040204} (\bibinfo {year} {2021})}\BibitemShut
	{NoStop}%
	\bibitem [{\citenamefont {Imamo\={g}lu}\ \emph {et~al.}(1999)\citenamefont
		{Imamo\={g}lu}, \citenamefont {Awschalom}, \citenamefont {Burkard},
		\citenamefont {DiVincenzo}, \citenamefont {Loss}, \citenamefont {Sherwin},\
		and\ \citenamefont {Small}}]{Imamoglu1999}%
	\BibitemOpen
	\bibfield  {author} {\bibinfo {author} {\bibfnamefont {A.}~\bibnamefont
			{Imamo\={g}lu}}, \bibinfo {author} {\bibfnamefont {D.~D.}\ \bibnamefont
			{Awschalom}}, \bibinfo {author} {\bibfnamefont {G.}~\bibnamefont {Burkard}},
		\bibinfo {author} {\bibfnamefont {D.~P.}\ \bibnamefont {DiVincenzo}},
		\bibinfo {author} {\bibfnamefont {D.}~\bibnamefont {Loss}}, \bibinfo {author}
		{\bibfnamefont {M.}~\bibnamefont {Sherwin}}, \ and\ \bibinfo {author}
		{\bibfnamefont {A.}~\bibnamefont {Small}},\ }\href {\doibase
		10.1103/PhysRevLett.83.4204} {\bibfield  {journal} {\bibinfo  {journal}
			{Phys. Rev. Lett.}\ }\textbf {\bibinfo {volume} {83}},\ \bibinfo {pages}
		{4204} (\bibinfo {year} {1999})}\BibitemShut {NoStop}%
	\bibitem [{\citenamefont {Benito}\ \emph {et~al.}(2019)\citenamefont {Benito},
		\citenamefont {Petta},\ and\ \citenamefont {Burkard}}]{Benito2019}%
	\BibitemOpen
	\bibfield  {author} {\bibinfo {author} {\bibfnamefont {M.}~\bibnamefont
			{Benito}}, \bibinfo {author} {\bibfnamefont {J.~R.}\ \bibnamefont {Petta}}, \
		and\ \bibinfo {author} {\bibfnamefont {G.}~\bibnamefont {Burkard}},\ }\href
	{\doibase 10.1103/PhysRevB.100.081412} {\bibfield  {journal} {\bibinfo
			{journal} {Phys. Rev. B}\ }\textbf {\bibinfo {volume} {100}},\ \bibinfo
		{pages} {081412R} (\bibinfo {year} {2019})}\BibitemShut {NoStop}%
	\bibitem [{\citenamefont {Blais}\ \emph {et~al.}(2004)\citenamefont {Blais},
		\citenamefont {Huang}, \citenamefont {Wallraff}, \citenamefont {Girvin},\
		and\ \citenamefont {Schoelkopf}}]{Blais2004}%
	\BibitemOpen
	\bibfield  {author} {\bibinfo {author} {\bibfnamefont {A.}~\bibnamefont
			{Blais}}, \bibinfo {author} {\bibfnamefont {R.-S.}\ \bibnamefont {Huang}},
		\bibinfo {author} {\bibfnamefont {A.}~\bibnamefont {Wallraff}}, \bibinfo
		{author} {\bibfnamefont {S.~M.}\ \bibnamefont {Girvin}}, \ and\ \bibinfo
		{author} {\bibfnamefont {R.~J.}\ \bibnamefont {Schoelkopf}},\ }\href
	{\doibase 10.1103/PhysRevA.69.062320} {\bibfield  {journal} {\bibinfo
			{journal} {Phys. Rev. A}\ }\textbf {\bibinfo {volume} {69}},\ \bibinfo
		{pages} {062320} (\bibinfo {year} {2004})}\BibitemShut {NoStop}%
	\bibitem [{\citenamefont {Wang}\ \emph {et~al.}(2010)\citenamefont {Wang},
		\citenamefont {Shao}, \citenamefont {Zhao}, \citenamefont {Zhang},\ and\
		\citenamefont {Yeon}}]{Wang2010}%
	\BibitemOpen
	\bibfield  {author} {\bibinfo {author} {\bibfnamefont {H.-F.}\ \bibnamefont
			{Wang}}, \bibinfo {author} {\bibfnamefont {X.-Q.}\ \bibnamefont {Shao}},
		\bibinfo {author} {\bibfnamefont {Y.-F.}\ \bibnamefont {Zhao}}, \bibinfo
		{author} {\bibfnamefont {S.}~\bibnamefont {Zhang}}, \ and\ \bibinfo {author}
		{\bibfnamefont {K.-H.}\ \bibnamefont {Yeon}},\ }\href {\doibase
		10.1364/JOSAB.27.000027} {\bibfield  {journal} {\bibinfo  {journal} {J. Opt.
				Soc. Am. B}\ }\textbf {\bibinfo {volume} {27}},\ \bibinfo {pages} {27}
		(\bibinfo {year} {2010})}\BibitemShut {NoStop}%
	\bibitem [{\citenamefont {Bartkowiak}\ and\ \citenamefont
		{Miranowicz}(2010)}]{Bartkowiak2010}%
	\BibitemOpen
	\bibfield  {author} {\bibinfo {author} {\bibfnamefont {M.}~\bibnamefont
			{Bartkowiak}}\ and\ \bibinfo {author} {\bibfnamefont {A.}~\bibnamefont
			{Miranowicz}},\ }\href {\doibase 10.1364/JOSAB.27.002369} {\bibfield
		{journal} {\bibinfo  {journal} {J. Opt. Soc. Am. B}\ }\textbf {\bibinfo
			{volume} {27}},\ \bibinfo {pages} {2369} (\bibinfo {year}
		{2010})}\BibitemShut {NoStop}%
	\bibitem [{\citenamefont {Godfrin}\ \emph {et~al.}(2018)\citenamefont
		{Godfrin}, \citenamefont {Ballou}, \citenamefont {Bonet}, \citenamefont
		{Ruben}, \citenamefont {Klyatskaya}, \citenamefont {Wernsdorfer},\ and\
		\citenamefont {Balestro}}]{Godfrin2018}%
	\BibitemOpen
	\bibfield  {author} {\bibinfo {author} {\bibfnamefont {C.}~\bibnamefont
			{Godfrin}}, \bibinfo {author} {\bibfnamefont {R.}~\bibnamefont {Ballou}},
		\bibinfo {author} {\bibfnamefont {E.}~\bibnamefont {Bonet}}, \bibinfo
		{author} {\bibfnamefont {M.}~\bibnamefont {Ruben}}, \bibinfo {author}
		{\bibfnamefont {S.}~\bibnamefont {Klyatskaya}}, \bibinfo {author}
		{\bibfnamefont {W.}~\bibnamefont {Wernsdorfer}}, \ and\ \bibinfo {author}
		{\bibfnamefont {F.}~\bibnamefont {Balestro}},\ }\href {\doibase
		10.1038/s41534-018-0101-3} {\bibfield  {journal} {\bibinfo  {journal} {npj
				Quantum Information}\ }\textbf {\bibinfo {volume} {4}},\ \bibinfo {pages}
		{53} (\bibinfo {year} {2018})}\BibitemShut {NoStop}%
	\bibitem [{\citenamefont {Rasmussen}\ and\ \citenamefont
		{Zinner}(2020)}]{Rasmussen2020b}%
	\BibitemOpen
	\bibfield  {author} {\bibinfo {author} {\bibfnamefont {S.~E.}\ \bibnamefont
			{Rasmussen}}\ and\ \bibinfo {author} {\bibfnamefont {N.~T.}\ \bibnamefont
			{Zinner}},\ }\href {\doibase 10.1103/PhysRevResearch.2.033097} {\bibfield
		{journal} {\bibinfo  {journal} {Phys. Rev. Research}\ }\textbf {\bibinfo
			{volume} {2}},\ \bibinfo {pages} {033097} (\bibinfo {year}
		{2020})}\BibitemShut {NoStop}%
	\bibitem [{\citenamefont {Kim}\ and\ \citenamefont {Choi}(2018)}]{Kim2018}%
	\BibitemOpen
	\bibfield  {author} {\bibinfo {author} {\bibfnamefont {T.}~\bibnamefont
			{Kim}}\ and\ \bibinfo {author} {\bibfnamefont {B.-S.}\ \bibnamefont {Choi}},\
	}\href {\doibase 10.1038/s41598-018-23764-x} {\bibfield  {journal} {\bibinfo
			{journal} {Scientific Reports}\ }\textbf {\bibinfo {volume} {8}},\ \bibinfo
		{pages} {5445} (\bibinfo {year} {2018})}\BibitemShut {NoStop}%
	\bibitem [{\citenamefont {Kingma}\ and\ \citenamefont {Ba}(2014)}]{Kingma2014}%
	\BibitemOpen
	\bibfield  {author} {\bibinfo {author} {\bibfnamefont {D.~P.}\ \bibnamefont
			{Kingma}}\ and\ \bibinfo {author} {\bibfnamefont {J.}~\bibnamefont {Ba}},\
	}\href@noop {} {\enquote {\bibinfo {title} {Adam: A method for stochastic
				optimization},}\ } (\bibinfo {year} {2014}),\ \bibinfo {note}
	{arxiv:1412.6980}\BibitemShut {NoStop}%
	\bibitem [{\citenamefont {Bahnsen}(2020)}]{Bahnsen2020}%
	\BibitemOpen
	\bibfield  {author} {\bibinfo {author} {\bibfnamefont {E.}~\bibnamefont
			{Bahnsen}},\ }\href@noop {} {\enquote {\bibinfo {title} {Native gate
				exploitation in four-qubit quantum circuit creation},}\ } (\bibinfo {year}
	{2020}),\ \bibinfo {note} {https://www.bahnsen.dev/masters}\BibitemShut
	{NoStop}%
\end{thebibliography}

%

\clearpage
\onecolumngrid
\appendix

\section{Alternative topologies for the QFT algorithm with the diamond gate}\label{sec:altTopo}
	
	There are several other ways to combine the diamond gate for the QFT algorithm. Here we discuss two other approaches which might lessen unwanted cross talk and frequency crowding.

	\begin{figure*}[h]
		\centering
		\includegraphics[width=\columnwidth]{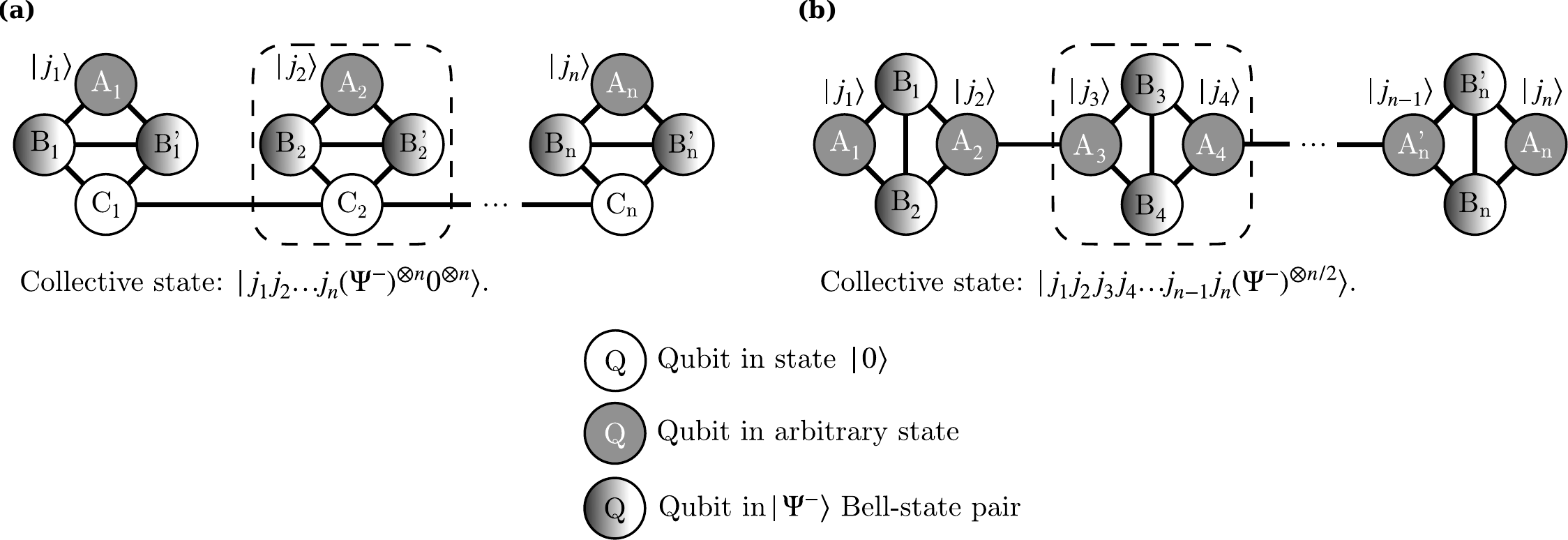}
		\caption{Alternative (to \cref{fig:qubit-concat}) arrangements of multiple diamond circuits interconnected to form a string to execution the diamond gate-assisted QFT algorithm.
		\textbf{(a)} The algorithm used for this can be seen in \cref{alg:dQFT2}. With this topology, we make use of the idle-state of the diamond, the CNS-gate, and the diamond-assisted method of performing a \Cr{}$_n$-gate of \cref{fig:URUflowChart}. \textbf{(b)} This configuration is very similar to that of \textbf{(a)}, except that we must be more careful when we use the CNS-gate to swap states. This uses two qubits for every input state, as opposed to four.}
		\label{fig:line_topology}
	\end{figure*}

	An example of such an alternative arrangement of the qubits is seen in \cref{fig:line_topology}(a). Here each diamond is connected via one of the target qubits. The control gates must be prepared in a $\ket{\Psi^-}$ state in order to exploit the idle state of the diamond gate. The non-connected target qubits, $A_i$, are prepared as the input states, while the connected qubits, $C_i$s, are prepared in the $\ket{0}$ state. Besides using the idle state of the diamond gate, we also employ the \cns gate. With all C-qubit states being $\ket{0}$, we can swap the states of each A- and C-qubit pair before this state moves on. Each time the state passes a diamond gate, we perform a \Cr{}$_n$ gate as described in \cref{sec:nativePhaseGate}. The algorithm used can be seen in \cref{alg:dQFT2}.

	{\centering
		\begin{minipage}{.95\linewidth}
			\begin{algorithm}[H]
				\caption{Native gate assisted QFT on string of diamonds [\cref{fig:line_topology}(b)]}
				\label{alg:dQFT2}
				\begin{algorithmic}[1]
					\algnewcommand\algorithmicto{\textbf{to}}
					\algrenewtext{For}[3]%
					{\algorithmicfor\ $#1 = #2$ \algorithmicto\ $#3$ \algorithmicdo}
					\State Initialize qubits A$_1$,\dots,A$_n$ with the states to transform ($\ket{j_1},\dots,\ket{j_n}$)
					\State Initialize qubits C$_1$,\dots,C$_n$ in state $\ket{0}$
					\State Initialize qubit pairs (B$_1$,B$_1'$),\dots,(B$_n$,B$_n'$) in state $\ket{\Psi^-}$
					\For{i}{1}{n}
					\State H on A$_i$
					\State Set qubits B$_i$ and B$_i'$ to $\ket{0}$
					\State Swap state of A$_i$ and C$_i$ with diamond CNS-gate
					\State Set qubit pair (B$_i$, B$_i'$) to $\ket{\Psi^-}$
					\For{j}{i+1}{n}
					\State \ciswap($t_g/2$) on (C$_{j-1}$, C$_j$)
					\State Set qubits B$_j$ and B$_j'$ to $\ket{0}$
					\State \ciswap($t_g/2$) on (A$_j$, B$_j$)
					\State R$_{j-i+1}$ on C$_j$ controlled by B$_j$ (using \cref{fig:URUflowChart})
					\State \ciswap($3t_g/2$) on (A$_j$, B$_j$)
					\State Set qubit pair (B$_j$, B$_j'$) to $\ket{\Psi^-}$
					\EndFor
					\State Swap state of C$_n$ back to C$_i$ with \ciswap($3t_g/2$)
					or correct for number of \ciswap($t_g/2$)'s along C-chain (mod 4) and readout state.
					\EndFor
					\State \text{Reverse order of C-qubits (these are the output qubits).}
				\end{algorithmic}
			\end{algorithm}
		\end{minipage}
		\par
	}
	\bigskip
	
	Another approach to the arrangement of the diamond gates can be seen in \cref{fig:line_topology}(b). In this case, only two qubits are used per input state, as opposed to the example in \cref{fig:line_topology}(a) where four qubits were used. Like the example in \cref{fig:line_topology}(a), it employs the idle state of the diamond, the \Cr$_n$ gate, and the \cns gate. Using this approach, we preserve the input state that is being transformed. The target of the \cns gate acts on the other qubit and not the one currently being transformed. This is similar to the \cns based pseudo-swap used in \cref{sec:CNS_QFT}.
	
	The algorithm associated with \cref{fig:line_topology}(b) is similar to the one in \cref{alg:dQFT2}. They both use three out of four control states of the diamond gate, $\ket{00}$, $\ket{11}$, and $\ket{\Psi^-}$. Both have the same circuit depth, $O(n^2)$, which can be reduced to $O(n \log n)$ with an approximate QFT approach.

\section{QFT algorithm for four qubits}\label{sec:QFTalgo}
\begin{figure}[h]
		\centering
		\includegraphics[scale=.7]{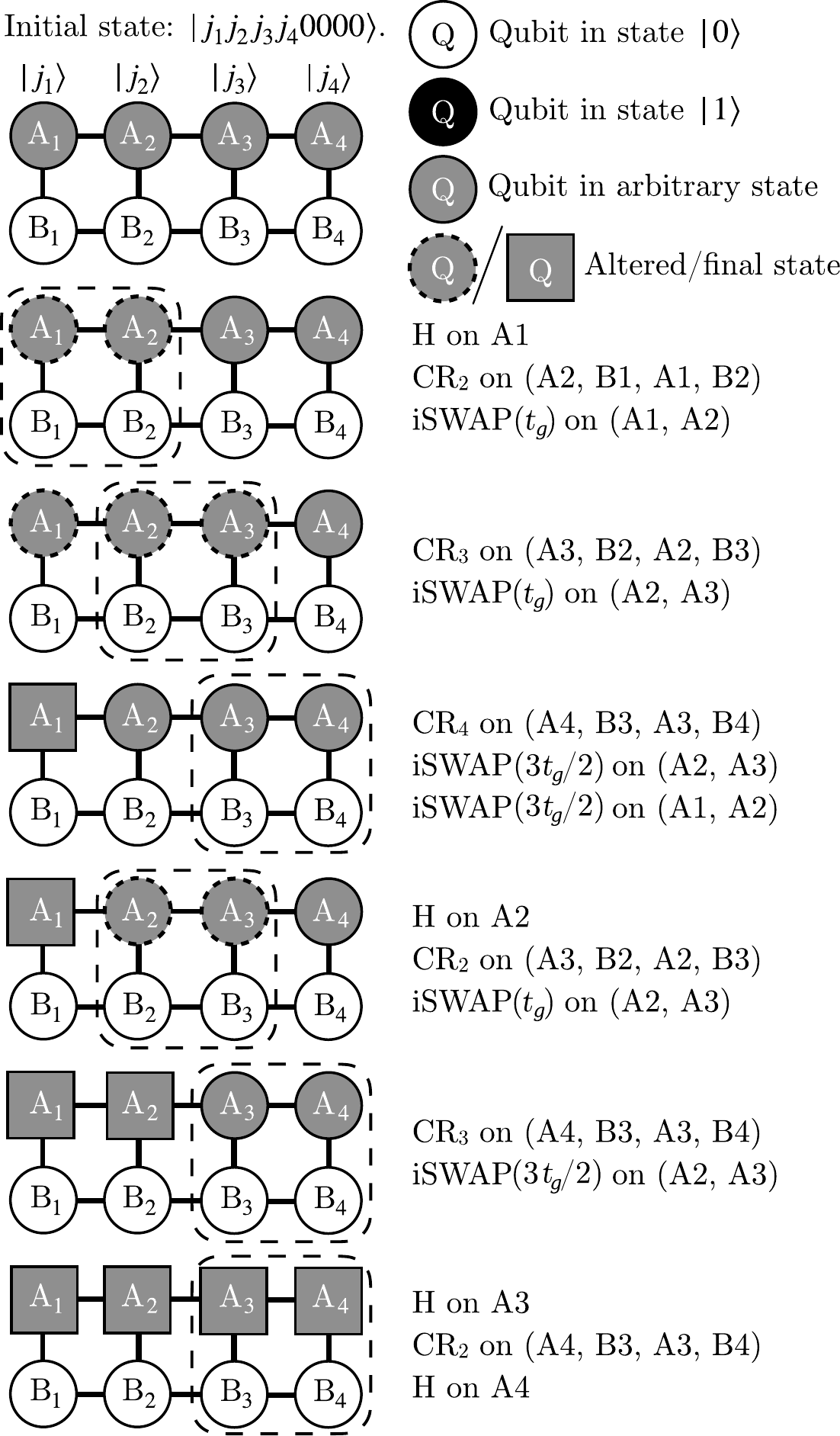}
		\caption{Example of the native gate-assisted QFT algorithm for input size $n=4$.
		The final cross-swap of A-qubits is not shown, such that the timing of the final states is highlighted.
		The dotted lines show the diamond part of the circuit in use for each step}
		\label{fig:qft4_alg}
	\end{figure}

	Here we present an example of the QFT algorithm for the four qubits in \cref{fig:qft4_alg}.
	The final cross-swap is not shown, such that it is clear when each qubit has reached their final state.
	Dotted lines shows the diamond circuit in use.
	
\end{document}